\documentclass[amsfonts,amsmath,prd,preprint,nofootinbib,a4paper]{revtex4}
\newcommand{\beq}{\begin{equation}}
\newcommand{\eeq}{\end{equation}}

\newcommand{\bsp}{\begin{split}}

\usepackage{epsfig,bbm,cancel,ulem}
\usepackage[breaklinks=true]{hyperref}
\usepackage{latexsym}
\usepackage[utf8]{inputenc}
\usepackage{amsmath}
\usepackage{amsfonts}
\usepackage{amssymb}
\usepackage{booktabs}
\usepackage{array}
\usepackage{tabularx}
\usepackage{multirow}
\usepackage{longtable}
\usepackage{ragged2e}
\usepackage{braket,booktabs}
\usepackage{braket,amsmath}
\usepackage{xcolor}
\usepackage{graphicx}

\newcommand{\ra}[1]{\renewcommand{\arraystretch}{#1}}
\begin{document}

\title{Gravitational Perturbation in Nonlocal Modified Tolman VII Model}

\author{Byon N. Jayawiguna}
\email{nugrahabyon312@gmail.com}
\author{Piyabut Burikham}
\email{piyabut@gmail.com}
\affiliation{High Energy Physics Theory Group, Department of Physics, Faculty of Science, Chulalongkorn University, Bangkok 10330, Thailand}
\def\changenote#1{\footnote{\bf #1}}

\begin{abstract}
In comparison to the original Tolman VII model, Exact Modified Tolman VII (EMTVII) with one additional parameter can increase the compactness of compact object. When the compactness is in the ultracompact regime, the quasinormal modes~(QNMs) of the trapped mode as well as the gravitational echoes become more viable. Starting with the EMTVII model, we introduce nonlocality into the matter sector and analyze the effective potential, the QNMs, and the gravitational echoes of the compact and ultracompact object in the nonlocal model. The nonlocal gravity version of EMTVII~(NEMTVII) is parametrized by the nonlocal parameter~($ \beta $), modified Tolman VII parameter ($ \alpha $), and the compactness ($ \mathcal{C}$). It is found that the nonlocal profile produces the smeared surface and consequently reduce the compactness. The maximum compactness $\mathcal{C}_{max}=0.4$ occurs when $\alpha=0=\beta$, i.e., EMTVII with no smearing. For relatively small value of $\beta = 0.01$ and the compactness $ \mathcal{C} \lesssim 0.2667$~(with $M=2.14$ solar masses, $R=11.835$ km at $\alpha=1.4$), the causality condition and the dominant energy condition~(DEC) are satisfied. The quasinormal modes of the gravitational perturbation are calculated using Bohr-Sommerfeld (BS) fitting and we find that the nonlocality produces less trapped modes than the original (EMTVII) counterpart. At high compactness, gravitational echoes are simulated numerically. Echoes are found to exist in the parameter space where the dominant energy condition and the causality condition are violated.

\end{abstract}

\maketitle
\thispagestyle{empty}
\setcounter{page}{1}

\section{Introduction}
A long journey from the first prediction up until the first direct detection of the black hole merger (GW150914) \cite{LIGOScientific:2016aoc}, the binary neutron stars (NS) inspiral (GW170817) \cite{LIGOScientific:2017vwq} and the two compact object coalesences (GW200105 and GW200115) recently in \cite{LIGOScientific:2021qlt} proves the existence of gravitational waves~(GW). At the dawn of the gravitational waves era, we still have rich information to obtain from observations which can be extensively studied both experimentally and theoretically. From the phenomenological point of view, NS and black holes could be just two specific objects of a larger family of the astrophysical objects with high compactness. There are numerous theoretical investigations to justify the existence of the horizonless compact object whose mass are large and the radius are relatively small: the black hole mimickers. These compact object are usually described by the compactness, $ \mathcal{C}\equiv GM/Rc^2, $ where $ G $ is the gravitational constant, $M$ and $R$ are the mass and the radius of the object and $c$ is the speed of light~(Here and henceforth, we will work in geometric unit where $G=c=1$). They can be classified as the following \cite{CP2019}:
\begin{enumerate}
	\item Compact object:~ $ \mathcal{C}>0.167, $
	\item Ultracompact Object: ~ $ \mathcal{C}>0.33, $
	\item Object above Buchdahl Limit:~ $ \mathcal{C}> 0.444, $
	\item Object with clean photon sphere:~ $ \mathcal{C}>0.49, $
	\item Object with near-horizon  scale quantum effects:~ $ \mathcal{C}\approx 0.5, $
	\item Black hole:~ $ \mathcal{C}=0.5. $
\end{enumerate}

NS with realistic equation of state (EoS) has a total mass between 1.4 and roughly two solar masses~($ M_{\odot} $) and its radius is roughly within the range of $10-20$ km~\cite{Clark:2002db,Antoniadis:2013pzd,Miller:2021qha}. Therefore, NS lies in the compact object category. If the object from the merger state is compact enough to have a photon sphere, then the post merger could have a sequence ringdown called the echoes that only emerge in the ultracompact region. However, some of the object exists, as a black hole mimicker, with the compactness larger than the Buchdahl limit. The object which satisfies the surface condition $r_{0}<2.038M$ allows the stable photon sphere to be formed~\cite{CP2019}. Examples of objects with clean photon sphere are the fuzzball~\cite{Hertog:2017vod,Bianchi:2020bxa,Bianchi:2020miz,Bah:2021jno,Bianchi:2021xpr,Bena:2022rna,Bianchi:2022qph} and gravastar~\cite{Mazur:2001fv,Chirenti:2007mk,Pani:2009hk,Pani:2009ss,Pani:2010em,Chirenti:2016hzd,Uchikata:2016qku,Volkel:2017ofl,Volkel:2017kfj}. If the quantum corrections are taken into account, the gravitational-wave signatures in the train-echo form could exist~\cite{Cardoso:2016oxy} for exotic compact object with $\mathcal{C}\approx 0.5$. Finally, the black hole's compactness is exactly at $\mathcal{C}=0.5$. 

Gravitational echoes provide information of the effective potential, the quasinormal modes~(QNMs) and the eigen solutions of the time dependent Regge-Wheeler equation (RW). The production of gravitational echoes (waves) requires the existence of photon sphere around the object, i.e., when the compactness reaches $ \mathcal{C}>1/3$.  It is intriguing to investigate the building blocks of matter which support this kind of compactness. Among the solutions to the Einstein field equations in general relativity, Tolman VII (TVII) \cite{Tolman:1939jz,Lattimer:2000nx,Jiang:2019vmf} model has a number of interesting physical properties. It has an analytic solution and represents a realistic NS which brings us the important information about the Equation of State (EoS) including nuclear physics~\cite{Raghoonundun:2015wga,Raghoonundun:2016cun,Hensh:2019rtb,Stuchlik:2021coc,Stuchlik:2021vdd,Posada:2021zxk,Posada:2023bnm}. For TVII, the scaling axial and polar GW modes are studied in Ref.~\cite{Tsui:2004qd} whereas the quasinormal modes are studied in Ref.~\cite{Neary:2001ai,Neary:2001yg}. Subsequently, Jiang and Yagi \cite{Jiang:2019vmf} introduce a modified Tolman-VII (MTVII): an additional quartic term to energy density profile with a free parameter, $ \alpha, $ in order to describe more realistic NS. The numerical solution of this model is studied by Ref.~\cite{Posada:2022lij} which we will call the exact numerical solution of the MTVII (EMTVII) in order to distinguish it from the original MTVII approximate solution studied in Ref.~\cite{Jiang:2019vmf}. It is found that the relevant range of the free parameter is $ \alpha\in[0,2]$, otherwise the EoS becomes unphysical due to negative density near the surface. The maximum compactness of the EMTVII model is $ \mathcal{C}_{max}=0.4 $ for $ \alpha=0 $ and the causal limit is $ \mathcal{C}=0.276 $ for $ \alpha=1.3$.

The post-merger state of the exotic compact object with $\mathcal{C}\approx 0.5$ is almost identical to a black hole. The ringdown waveform from the coalescence of the binary objects provides a tool for testing the existence of the compact object~(horizonless) and BH dominated by QNMs. The quantum effects on the near-horizon scale could appear in the late-time ringdown waveform~\cite{Cardoso:2016oxy,Cardoso:2016rao,Conklin:2017lwb,Oshita:2018fqu,Wang:2019rcf,Cardoso:2019apo,Buoninfante:2020tfb}. In Ref.~\cite{Maggio:2020jml}, the pathology in the classical black hole is replaced by the horizonless, singularity-free solution, by extending the black hole concepts into the membrane paradigm. Recent observations can be used to put constraint on the compactness in the order of $\mathcal{C}>0.49 $ (99\% of the black hole compactness). 

Another black hole mimicker is the nonlocal star which can smear out the curvature singularity via the nonlocal interaction that could potentially regulate the UV behavior of nonrenormalizable gravitational theory~\cite{Nicolini:2005vd,Koshelev:2017bxd,Buoninfante:2018rlq,Buoninfante:2018xif,Biswas:2011ar,Nicolini:2012eu,Frolov:2015bta,Buoninfante:2019swn,Buoninfante:2019zws}. Interestingly, nonlocality could amplify the gravitational echoes in certain system~\cite{Buoninfante:2019teo}. The other types of nonlocal theory such as the generalized uncertainty principle (GUP) \cite{Battisti:2007jd,Shibusa:2007ju,Battisti:2007zg,Nozari:2009cs,Bojowald:2011jd,Huang:2012hia,Sprenger:2012uc,Isi:2013cxa,Ali:2013ma,Tawfik:2014zca,Bruneton:2016yws,Li:2020vqg} and noncommutative geometry \cite{Nicolini:2005zi,Nicolini:2005de,Rizzo:2006zb,Ansoldi:2006vg,Casadio:2008qy,Spallucci:2008ez,Banerjee:2009xx,Gingrich:2010ed,Modesto:2010rv,Mann:2011mm,Mureika:2011py,Nicolini:2005vd,Nicolini:2009gw,Nicolini:2011dp,Nicolini:2008aj,Banerjee:2008gc} are also discussed.

In this work, we investigate the impact of nonlocal gravity to the MTVII density model by analytical and numerical analyses. We focus on the ultracompact object in nonlocal gravity with ideal and isotropic fluid. It is found that the ultracompact nonlocal object could exist in the range $ \beta\in[0,~0.15]. $ The effective potential, quasinormal modes, and the eigen solutions of Regge-Wheeler~(RW) equation are subsequently obtained and explored.

This work is organized as the following. In Section \ref{SectI}, we review the ordinary and modified Tolman VII model and their limitation. In Section \ref{SectII}, nonlocal gravity coupled with the MTVII model is considered and the equations of motion are solved numerically. Since the maximum compactness and the energy condition depend on the choices of MTVII parameter $\alpha $, nonlocal parameter $\beta$, and the compactness $\mathcal{C}$, the results are discussed with respect to those parameters in Section \ref{SectIII}. In Section \ref{SectIV}, we employ axial perturbations to the metric describing nonlocal gravity coupled with the EMTVII. The effective potential, quasinormal modes, and the gravitational echoes are examined. Section \ref{SectV} concludes our work.


\section{(Exact) Modified Tolman VII Model}  \label{SectI}

In this section, we will briefly review the ordinary TVII model and the (modified) MTVII model. Generically, the spherically symmetric metric is in the form
\begin{equation}
\label{ansatz}
ds^2 = - e^{\nu(r)} dt^2 + e^{\lambda(r)} dr^2 + r^2 d\Omega^2,
\end{equation}
where $ tt $ and $ rr $ components are a function of $ r, $. The matter sector for this case is the perfect fluid tensor energy momentum which can be written as
\begin{equation}
\label{matter}
T_{\mu\nu} = (\rho + p) u_{\mu}u_{\nu} + p g_{\mu\nu},
\end{equation}
where $ \rho, p, u_{\mu} $ are energy density, pressure, and 4-velocity respectively. The Einstein field equation, $ G_{\mu\nu}=8\pi T_{\mu\nu}, $ gives
\begin{eqnarray}
&& \label{meq} m'= 4\pi r^2 \rho, \\ && \label{nueq} \left(\frac{e^{-\lambda}-1}{r^2}\right)' + \left(\frac{e^{-\lambda}\nu'}{2r}\right)' + e^{-(\lambda+\nu)} \left(\frac{e^{\nu}\nu'}{2r}\right)'=0, \\ && \label{peq} e^{-\lambda}\left(\frac{\nu'}{r}+\frac{1}{r^2}\right)-\frac{1}{r^2}= 8\pi p,
\end{eqnarray}
Note that the prime symbol $ ' $ denotes $ d/dr $ and $ e^{-\lambda}\equiv 1-2m(r)/r $. The first eqn.~(\ref{meq}) represents the equation for the mass of the star. The second eqn.~(\ref{nueq}) represents the equation for solving $ e^{\nu} $, and the pressure can be obtained by substituting the $ \nu(r) $ solution to the third eqn.~(\ref{peq}). In the Tolman VII model, the corresponding energy density profile is defined in a simple quadratic form which can be written as
\begin{equation}
\rho (r) \equiv \rho_{c}\left[1-(r/R)^2\right], 
\end{equation} 
where $ \rho_{c} $ is the energy density at the center and $ R $ is a stellar radius. From these equations, we can obtain a complete solution of the metric inside the star within the TVII model. Substitute the density profile into Eqn. \eqref{meq} and integrate with the appropriate choices of boundary conditions at $r=R$ gives the mass profile of this model
\begin{eqnarray}
m_{\rm TVII}(r) = 4\pi \rho_{c} \left(\frac{r^3}{3}-\frac{r^5}{5R^2}\right),
\end{eqnarray}
hence, the explicit value of $ \rho_{c} $ can be obtained by using condition $ M \equiv m(R) $. The result is $ \rho_{c} = 15 M/8\pi R^3 $. Then, we can obtain the metric in the $ rr $-component as
\begin{eqnarray}
e^{-\lambda_{\rm TVII}} (r) = 1-\frac{M}{ R}\left(\frac{5r^2}{R^2}-\frac{3r^4}{R^4}\right).
\end{eqnarray}
This result can be used to obtain the $ \nu $-function by substituting the latter equation into eqn. \eqref{nueq}. The result reads
\begin{equation}
e^{\nu_{\rm TVII}} (r) = C_{1,~{\rm TVII}} \cos^{2}[\phi_{\rm TVII} (r)],
\end{equation}
where 
\begin{equation}
\phi_{\rm TVII} (r)= C_{2,~{\rm TVII}} - \frac{1}{2}\log \bigg|\frac{r^2}{R^2}-\frac{5}{6} +\sqrt{\frac{ e^{-\lambda_{\rm TVII}}}{3\mathcal{C}}}   \bigg|.
\end{equation}
Moreover, the pressure reads
\begin{equation}
p = \frac{1}{4\pi R^2} \left[ \sqrt{3\mathcal{C} e^{-\lambda_{\rm TVII}}}\tan\phi_{\rm TVII}-\frac{\mathcal{C}}{2} \left(5-\frac{3r^2}{R^2}\right) \right].
\end{equation}
The $ C_{1, ~{\rm TVII}} $ and $ C_{2, ~{\rm TVII}} $ are defined as
\begin{eqnarray}
C_{1, ~{\rm TVII}} &\equiv& 1-\frac{5\mathcal{C}}{3}, \\ C_{2,~{\rm TVII}}&\equiv& \tan^{-1}\left[\sqrt{\frac{\mathcal{C}}{3(1-2\mathcal{C})}} ~~\right] + \frac{1}{2}\log\bigg|\frac{1}{6}+\sqrt{\frac{1-2\mathcal{C}}{3\mathcal{C}}}\bigg|,\\
\end{eqnarray}
where the compactness $ \mathcal{C}\equiv M / R. $ These integration constants are determined from the boundary condition at $ r=R $; by matching the exterior solution and use zero pressure value at the object's surface. The allowed compactness based on the finiteness of the pressure and the metric for the TVII model \cite{Jiang:2019vmf,Posada:2021zxk} is $ \mathcal{C} \lesssim 0.386$ whereas the allowed compactness with the causality condition  (subluminal sound speed, $ dp/d\rho<1 $) is $ \mathcal{C} \lesssim 0.26$. However, Jiang and Yagi \cite{Jiang:2019vmf} proposed a new model by adding quartic term to the density model in order to describe more realistic NS. The density model reads
\begin{equation}
\label{rhomod}
\rho_{\rm mod}(r) = \rho_{c} \left[ 1-\alpha \frac{r^2}{R^2} + (\alpha-1) \frac{r^4}{R^4}   \right]
\end{equation}
with a new parameter $ \alpha $. The central density, $ \rho_{c}, $ can be written as
\begin{equation}
\rho_{c} = \frac{105 M}{8\pi R^3(10-3\alpha),}
\end{equation}
The mass and the $g^{rr}$ metric component are then given by
\begin{eqnarray}
m_{\rm mod}(r) &=& \frac{M}{ (20-6\alpha)} \left[35\frac{r^3}{R^3} - 21\alpha\frac{r^5}{R^5} + 15 (\alpha-1) \frac{r^7}{R^7}   \right],\nonumber \\ e^{-\lambda_{\rm mod}} &=& 1-\frac{M}{ R(10-3\alpha)} \left[35\frac{r^2}{R^2} - 21\alpha\frac{r^4}{R^4} + 15 (\alpha-1) \frac{r^6}{R^6}    \right],
\end{eqnarray}
where ``mod" denotes the modified solution corresponds to the density in \eqref{rhomod}. In MTVII model with additional term in the density presented in \eqref{rhomod}, the $\nu(r)$ equation is no longer analytic. Jiang and Yagi make use of approximate solution by using $ g_{rr} $ metric from the original Tolman VII with the modified boundary condition~\cite{Jiang:2019vmf}. Later the equations are numerically solved by Camilo Posada, \textit{et al}~\cite{Posada:2022lij} and the physically allowed range is in $\alpha \in [0,2]$. The allowed (maximum) compactness for compact object in EMTVII, as presented in \cite{Posada:2022lij}, is $ \mathcal{C}\leq 0.4 $ for $ \alpha=0 $, whereas in causality domain we have $ \mathcal{C} \leq 0.276 $ for $ \alpha=1.3 $. Notably, both upper bounds on the compactness are greater than the TVII model. As a further extension, we will include the nonlocality effect to the matter sector with EMTVII model and consider how nonlocal parameter, $ \beta, $ affect the range of compactness $ \mathcal{C}. $ We will also compare our solutions with the EMTVII.

\section{Nonlocal Gravity with (Exact)~Modified Tolman VII Model (NEMTVII)}   \label{SectII}

The nonlocality in the gravity sector can be mapped into the nonlocality in the matter sector and has been used recently to study the impact of nonlocality in a black hole~\cite{Nicolini:2012eu,Gaete:2010sp,Mureika:2010je,Moffat:2010bh,Modesto:2010uh}. The action is
\begin{equation}
\label{action}
S = \frac{1}{16\pi } \int d^{4}x \sqrt{-g} ~\mathcal{R}(x) + S_{\textrm{matter}},
\end{equation}
where the ordinary Ricci tensor, $ R, $ is embedded through the following form
\begin{equation}
\mathcal{R}(x)= \int d^{4}y \sqrt{-g} \mathcal{A}^2(x-y)R(y),
\end{equation}
with the bi-local distribution defined by
\begin{equation}
\mathcal{A}^2(x-y)\equiv \mathcal{A}^2(\Box_{x}) \delta^{(4)}(x-y).
\end{equation}
The operator represents nonlocal interaction between 2 points in spacetime and could potentially correlate points in the inner and outer region of the black hole horizon.

Here, the operator $\Box_{x}= \beta g_{\mu\nu}\nabla^{\mu}\nabla^{\nu}, $ is dimensionless D'Alembertian operator. We can obtained the Einstein field equation~(EFE) by varying the action in Eqn.~\eqref{action} with respect to the metric tensor $g_{\mu\nu}$ and neglecting the surface term related to variation of the D'Alembertian. The nonlocal EFE reads
\begin{equation}
\label{efeng}
\mathcal{A}^2(\Box) \left(R_{\mu\nu}-\frac{1}{2}g_{\mu\nu}R\right) = 8\pi T_{\mu\nu}.
\end{equation}
The bi-local operator works on the gravity sector. However, the EFE can be written in the following form
\begin{eqnarray}
\label{efeng2}
R_{\mu\nu}-\frac{1}{2}g_{\mu\nu}R = 8\pi  \mathcal{T}_{\mu\nu},
\end{eqnarray}
where $\mathcal{T}_{\mu\nu}\equiv \mathcal{A}^{-2}(\Box)T_{\mu\nu} $. The operator $ \mathcal{A} $ has a unique and well-defined inverse \cite{Nicolini:2012eu}. The role of the nonlocal operator now is to correlate energy-matter between 2 points in spacetime. The action \eqref{action} can be represented in two possible ways; Eqn. \eqref{efeng} denotes the nonlocal geometry coupled to the ordinary matter, whereas Eqn. \eqref{efeng2} denotes the ordinary local gravity coupled to the generalized matter. The two forms are equivalent \cite{Nicolini:2012eu}. In the entire calculation in this work, we use Eqn.~\eqref{efeng2} as the representation of nonlocal EFE to be solved. The nonlocal effect appears in the matter sector, and hence, the tensor energy-momentum presented in Eqn.~\eqref{matter} is modified to
\begin{equation}
\mathcal{T}_{\mu\nu} = (\tilde{\rho}+\tilde{p})u_{\mu}u_{\nu} + \tilde{p} ~ g_{\mu\nu},
\end{equation}
and the component of the Einstein field equation reads
\begin{eqnarray}
&& \label{00} \tilde{m}'= 4\pi  r^2 \tilde{\rho}, \\ && \label{22} \left(\frac{e^{-\lambda}-1}{r^2}\right)' + \left(\frac{e^{-\lambda}\nu'}{2r}\right)' + e^{-(\lambda+\nu)} \left(\frac{e^{\nu}\nu'}{2r}\right)'= 0, \\ && \label{11} e^{-\lambda}\left(\frac{\nu'}{r}+\frac{1}{r^2}\right)-\frac{1}{r^2}= 8\pi \tilde{p},
\end{eqnarray}
The EFEs for the isotropic perfect fluid lead to three equations for the matter variables, $\tilde{\rho}~ \textrm{and}~ \tilde{p} $, and the two metric variables, $\nu(r) ~\textrm{and}~ \lambda(r) $. Our next task is to obtain the modified nonlocal density profile. In order to find $\tilde{\rho}, $ it is necessary to choose a particular $ \mathcal{A}(\Box)$. To the best of our knowledge, there is no experimental information about the quantum gravity hence there is no restrictions for choosing a suitable form of $\mathcal{A}(\Box)$~\cite{Isi:2013cxa}. We choose the function so that the higher momenta can be taken into account
\begin{equation}
\mathcal{A}(\Box)=(1-\Box)^{1/2}.
\end{equation}   						
By using the Schwinger representation\footnote{The definition can be written as \cite{Isi:2013cxa} $\hat{\Delta}^{\gamma}=\frac{1}{\Gamma(-\gamma)} \int_{0}^{\infty} ds~s^{-\gamma-1}e^{-s\hat{\Delta}}.$ Since the operator acting on the density is $ \mathcal{A}^{-2} $, hence $ \hat{\Delta}=(1-\Box) $ and $ \gamma=-1. $}, we obtain 
\begin{equation}
\mathcal{A}^{-2} = \int_{0}^{\infty}ds~e^{-s[1+(-\Box)]}.
\end{equation}	    
Hence the density can be written as
\begin{eqnarray}
\label{rhong}
\tilde{\rho}= \mathcal{A}^{-2} \rho &=& \frac{1}{(2\pi)^{3}} \int_{0}^{\infty} ds~ e^{-s(1-\beta \nabla^2)} \int d^{3}p~ \rho(p)~ e^{i\vec{x}.\vec{p}}  ,\\ &=& \frac{1}{(2\pi)^{3}} \int_{0}^{\infty} ds~ e^{-s(1+\beta p^2)} \int d^{3}p~ \rho(p)~ e^{i\vec{x}.\vec{p}},\\ &=& \label{rong} \frac{1}{2\sqrt{\beta}x} \int_{0}^{R} dx'~x'~\rho(x') \left[ e^{-\frac{|x-x'|}{\sqrt{\beta}}}- e^{-\frac{|x+x'|}{\sqrt{\beta}}}  \right],
\end{eqnarray}
where $\beta$ is the nonlocal parameter that has a dimension $\textit{length}^2 $, $ \beta \sim \ell^2 $. After some calculation, we obtain the expression for the nonlocal density. Since we have absolute signs, this expression can be split into interior and exterior region and do the integration with appropriate interval. In general, we can put any density $\rho(x')$ in the integrand. In this work we only use the (E)MTVII density. Changing variables and inserting the model into Eqn.~\eqref{rong} gives
\begin{eqnarray}
\label{roint}
\tilde{\rho}_{\textrm{int}} (r) &=& \frac{\rho_{c}~e^{-(r+R)/\sqrt{\beta}}}{r/\sqrt{\beta}} \bigg\lbrace \left(1-e^{2r/\beta}\right) \bigg[ (\alpha-2)+(7\alpha-10)\frac{\sqrt{\beta}}{R} +3(9\alpha-10)\frac{\beta}{R^2} \nonumber \\ && +60(\alpha-1)\frac{\beta^{3/2}}{R^3}+60(\alpha-1)\frac{\beta^2}{R^4} \bigg] + \frac{e^{(r+R)/\sqrt{\beta}}~r}{\sqrt{\beta}}  \bigg[1-\alpha \frac{r^2}{R^2}+(\alpha-1) \frac{r^4}{R^4}\nonumber \\ && -6\alpha \frac{\beta}{R^2}+120(\alpha-1) \frac{\beta^2}{R^4}+20(\alpha-1)\frac{\beta}{R^2}   \bigg]\bigg\rbrace,
\end{eqnarray}
and
\begin{eqnarray}
\label{roext}
\tilde{\rho}_{\textrm{ext}} (r)&=& \frac{\rho_{c}~ e^{-\frac{(r+R)}{\sqrt{\beta}}}}{r/\sqrt{\beta}} \bigg\lbrace \left(1-e^{2R/\sqrt{\beta}}\right)\bigg[(\alpha-2) +3(9\alpha-10)\frac{\beta}{R^2} +60(\alpha-1)\frac{\beta^2}{R^4}  \bigg]\nonumber \\ &&+\left(1+e^{2R/\sqrt{\beta}}\right) \bigg[(7\alpha-10)\frac{\sqrt{\beta}}{R}+60 (\alpha-1)\frac{\beta^{3/2}}{R^3} \bigg]   \bigg\rbrace.
\end{eqnarray}
The modified nonlocal density profile is characterized by its star radius $ R, $ nonlocal parameter $ \beta $ and the energy density value at the center. The mass can be written as 
\begin{eqnarray}
\label{massint}
\tilde{m}_{\textrm{int}} (r) &=& 4\pi\rho_{c}\bigg\lbrace \frac{(\alpha-1)}{7}\frac{r^7}{R^4} +\frac{r^5}{R^2}\left[\frac{4\beta}{R^2}(\alpha-1)-\frac{\alpha}{5}\right] + \frac{r^3}{3}\left[1-\frac{6\alpha\beta}{R^2}+\frac{120 (\alpha-1)\beta^2}{R^4}\right]\nonumber\\&&-\left[e^{\frac{r-R}{\sqrt{\beta}}}(\beta r-\beta^{3/2}) + e^{-\frac{r+R}{\sqrt{\beta}}}(\beta r+\beta^{3/2}) \right]\times\bigg[(\alpha-2)+(7\alpha-10)\frac{\sqrt{\beta}}{R} \nonumber \\ && +\frac{3\beta}{R^2}(9\alpha-10)+\frac{60\beta^{3/2}(\alpha-1)}{R^3}+\frac{60\beta^2(\alpha-1)}{R^4} \bigg] \bigg\rbrace,
\end{eqnarray}
and
\begin{eqnarray}
\label{massext}
\tilde{m}_{\textrm{ext}} (r)&=& 4\pi \rho_{c} \left[e^{-\frac{(r+R)}{\sqrt{\beta}}} (\beta r+\beta^{3/2}) - e^{-\frac{2R}{\sqrt{\beta}}} (\beta R + \beta^{3/2})\right]\times\bigg\lbrace \left(e^{2R/\sqrt{\beta}}-1\right)\bigg[(\alpha-2) \nonumber \\ &&+3(9\alpha-10)\frac{\beta}{R^2} +60(\alpha-1)\frac{\beta^2}{R^4}  \bigg]-\left(1+e^{2R/\sqrt{\beta}}\right) \bigg[(7\alpha-10)\frac{\sqrt{\beta}}{R}+60 (\alpha-1)\frac{\beta^{3/2}}{R^3} \bigg] \bigg\rbrace.\nonumber\\
\end{eqnarray}
We can subsequently solve the second-order differential equation for $\nu(r)$ and obtain the nonlocal pressure. In the next section, we will perform the numerical calculation and analyze the parametric plot between the maximum compactness and nonlocal parameter for each value of $\alpha$.

\section{Validity of the NEMTVII Solution}  \label{SectIII}

\begin{figure}[h!]
	\centering
		\includegraphics[width=0.49\linewidth]{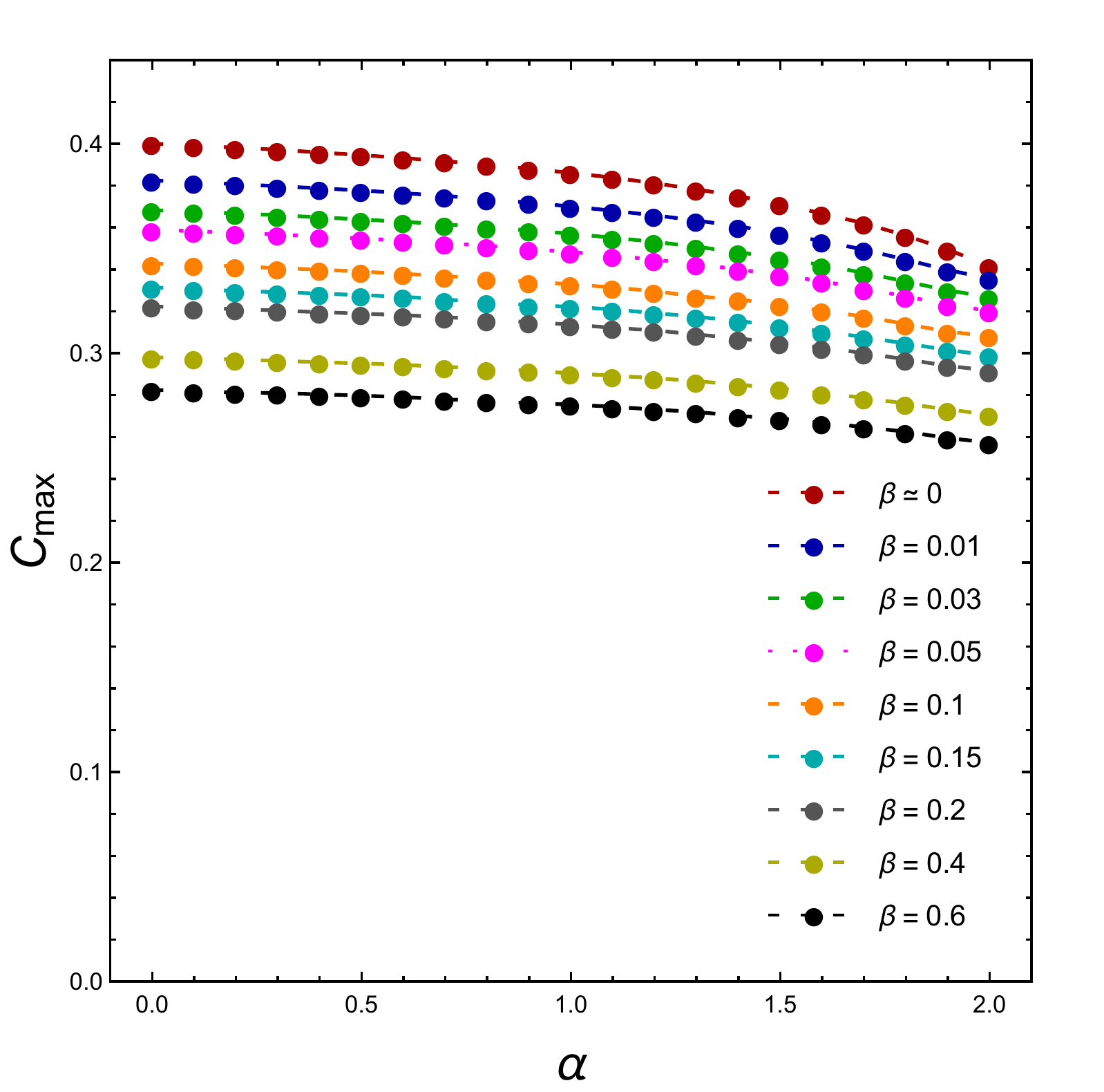}
	\includegraphics[width=0.49\linewidth]{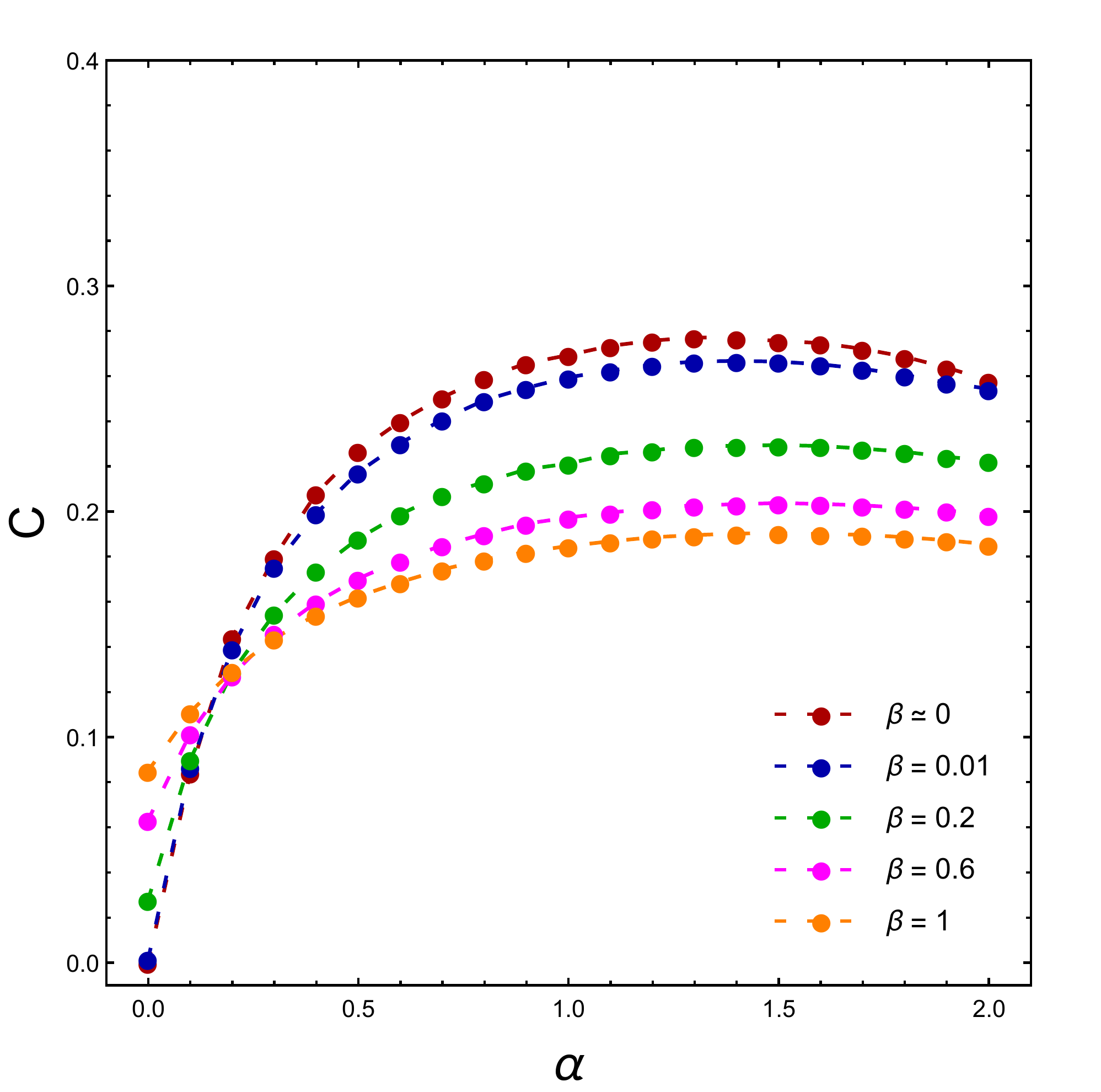}
	\caption{[Left] Maximum compactness versus $\alpha$ for $\alpha \in [0,2]$. [Right] Parametric plot between compactness $\mathcal{C}$ within the causal limit ($ c_{s}\big|_{r=0} \leq 1 $) and $ \alpha $. Several exact values are presented in \ref{AppA} and \ref{AppB}.}
	\label{fig:1}
\end{figure}

In Section~\ref{SectI} and Ref.~\cite{Posada:2022lij}\footnote{Orange solid line in FIG.6}, it was observed that the allowed maximum compactness predicted by EMTVII model for $ 0\leq\alpha<1~(1<\alpha\leq2)$ is higher~(lower) than the TVII model $ (\alpha=1) $. When the nonlocality is included in the matter sector, the nonlocal parameter should affect compactness as well as radius of the star. In this section, the Einstein's equations are solved numerically under suitable boundary conditions that match the Schwarzschild solution. From the numerical solutions, we examine the parameter space of the physically allowed solution as shown in Fig.~\ref{fig:1}.

In the left panel of Fig.~\ref{fig:1}, $ \mathcal{C}_{max} $ represents the maximum compactness for each model. The right panel of Fig.~\ref{fig:1} represents the parametric plot between compactness $\mathcal{C}$ and $ \alpha $. The red dotted line in the left~(right) figures represent the maximum compactness without~(within) the causal limit, $ c_{s}\big|_{r=0} \leq 1$, in EMTVII model \cite{Posada:2022lij}. Incorporating nonlocality makes the compactness decreases due to larger radius of the smeared surface of the star. For each $ \beta, $ the maximum compactness only occurs when $ \alpha=0 $. Nonlocal parameter, $ \beta, $ covers the ultracompact object only in the region $ \beta\in[0,~0.15]. $ The exact values for the maximum compactness in the nonlocal gravity model that we will use in the subsequent section are $ \mathcal{C}_{max}=0.382 $ for $ \beta=0.01 $ and $ \mathcal{C}_{max}=0.368 $ for $ \beta=0.03 $, both at $ \alpha=0$. On the other hand within the causal limit, the nonlocality can increase the compactness only for certain values of $ \alpha $. The maximum compactness at causal limit is $\mathcal{C}=0.2667$ for $\beta = 0.01$ and $\alpha=1.4 $. Note that the maximum compactness in red dotted line (EMTVII) in both panel of Fig.~\ref{fig:1} is the same as reported in Ref.~\cite{Posada:2022lij} (solid orange and blue line in Fig. 6) for the exact numerical solution of MTVII whereas Ref.~\cite{Posada:2021zxk} uses approximate solution of MTVII.

\begin{figure}[h!]
	\centering
	\includegraphics[width=0.45\linewidth]{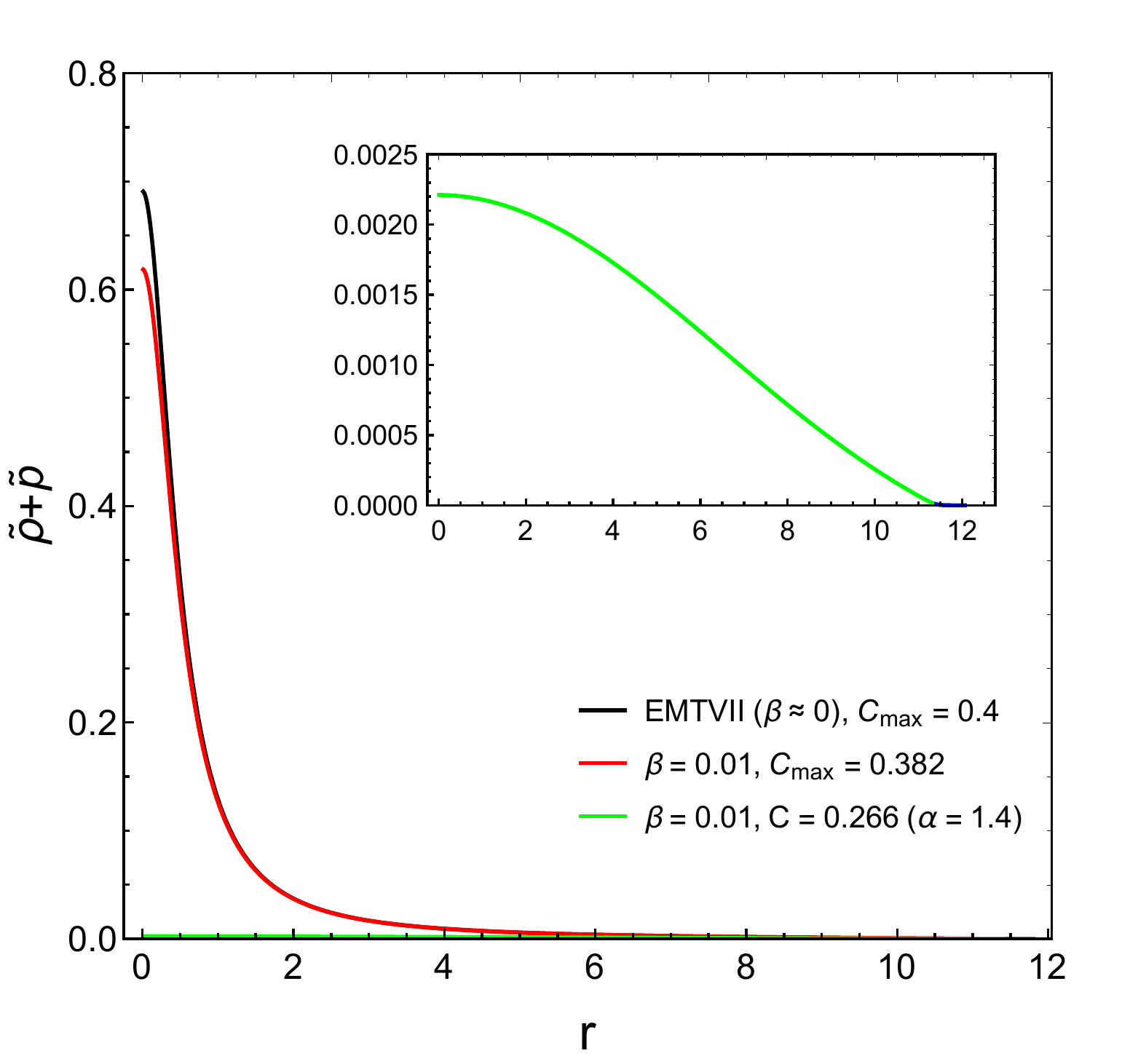}
	\includegraphics[width=0.45\linewidth]{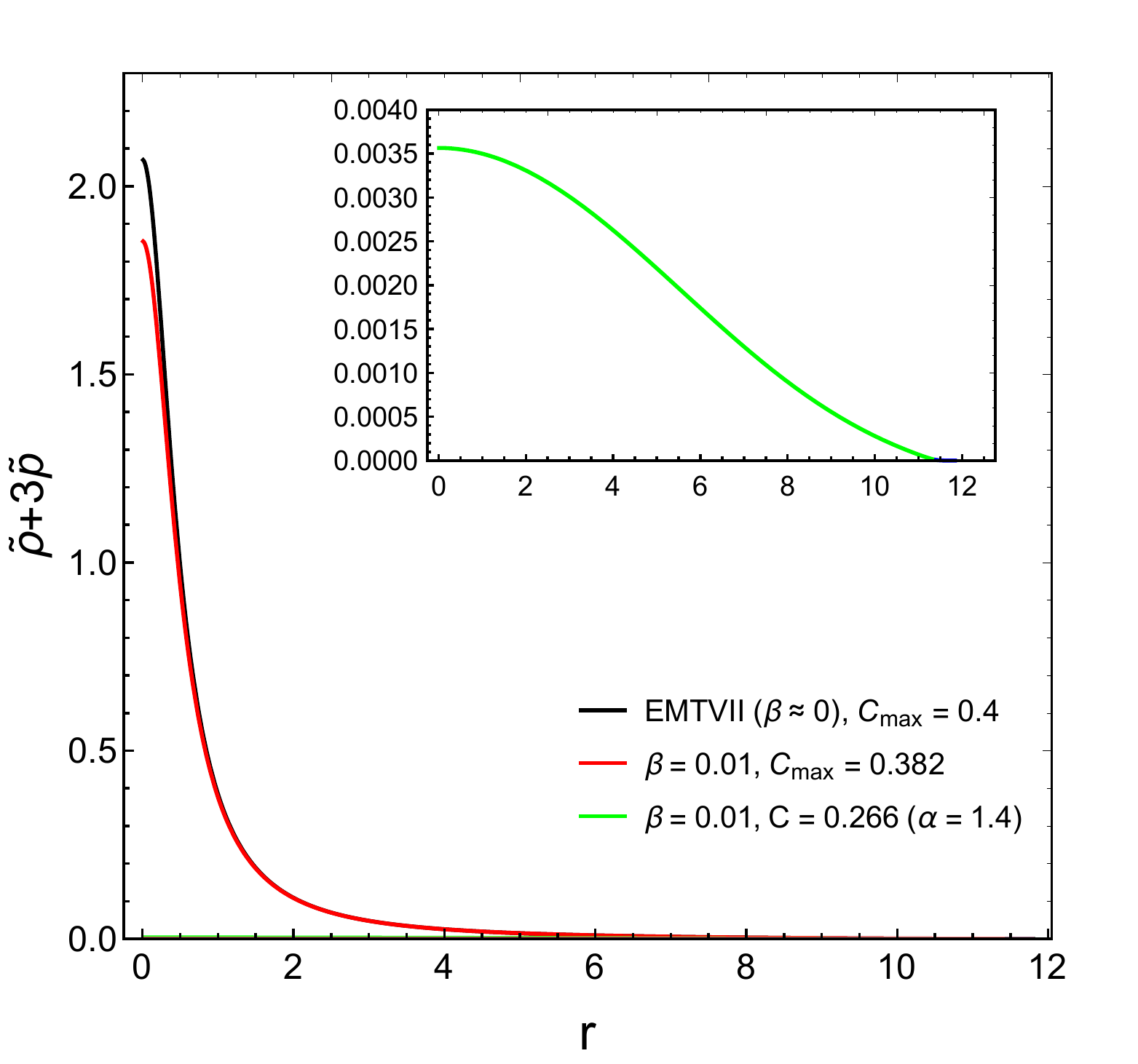}
	\includegraphics[width=0.45\linewidth]{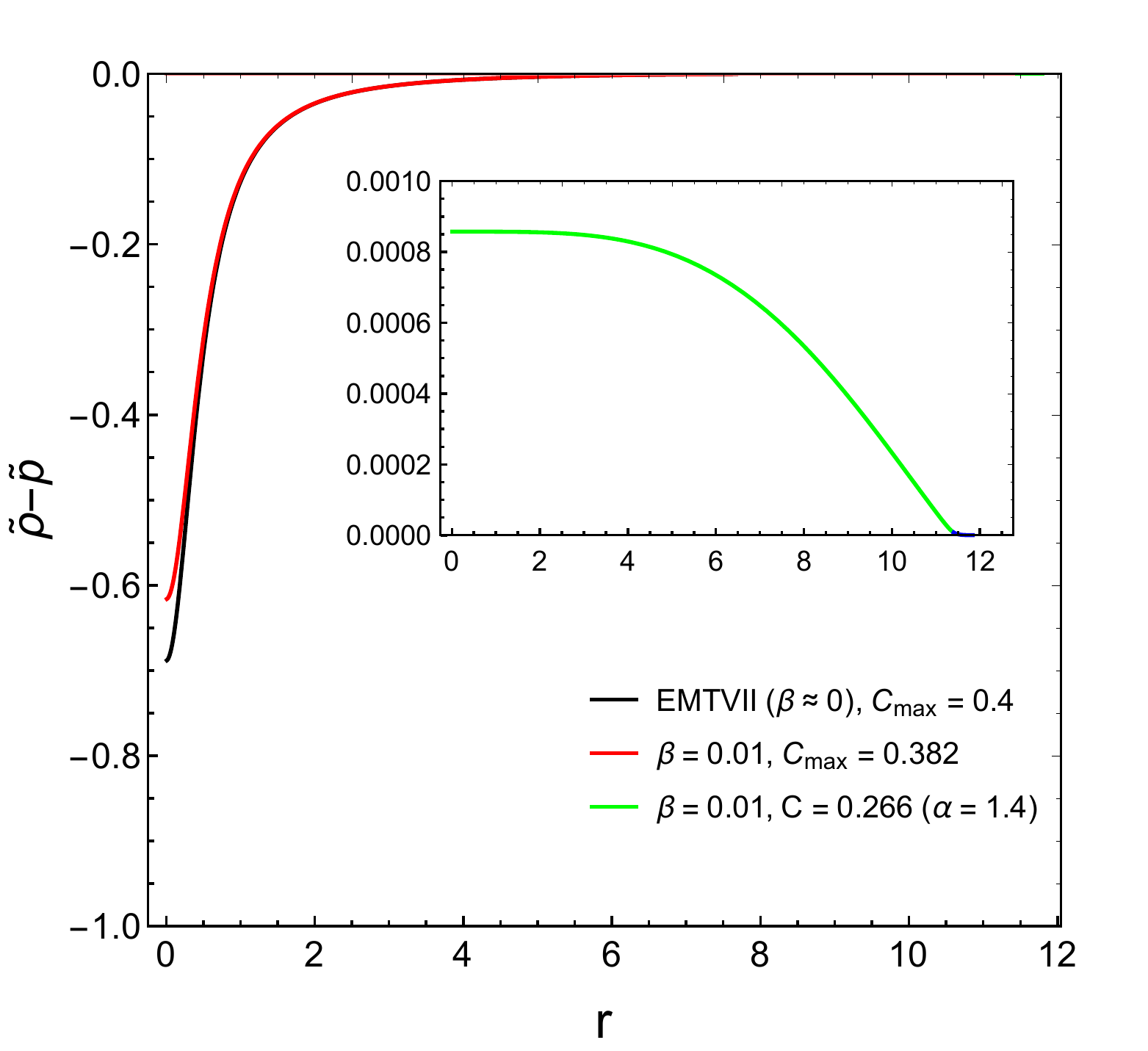}
	\caption{Profiles of the energy condition at maximum compactness with and without the causality condition. The null energy condition is automatically satisfied since it has non-negative density.}
	\label{fig:2}
\end{figure}

The physically acceptable density and the pressure profile should satisfy the energy conditions. Hence, it is crucial to check whether the equation of state of ultracompact stars within the NEMTVII model satisfies the energy condition. The energy conditions can be categorized into four types, i.e., Strong Energy Condition (SEC), Weak Energy Condition (WEC), Dominant Energy Condition (DEC), and Null Energy Condition (NEC). These conditions can be mathematically summarized as
\begin{eqnarray}
\textrm{SEC}&:&~~\tilde{\rho}+ \tilde{p} \geq 0,~~~\tilde{\rho}+3 \tilde{p} \geq 0,\\
\textrm{DEC}&:&~~\tilde{\rho} \geq |\tilde{p}|,\\
\textrm{WEC}&:&~~\tilde{\rho} \geq 0,~~~\tilde{\rho}+\tilde{p} \geq 0,\\
\textrm{NEC}&:&~~\tilde{\rho} \geq 0.
\end{eqnarray}
We also assume that matter is isotropic. In Fig.~\ref{fig:2}, we present corresponding combination between nonlocal density and nonlocal pressure relevant to the energy conditions. The black and red solid line denote the maximum compactness the star can have as predicted by EMTVII and the nonlocal extension model whereas the green line denotes the compactness that obeys causal condition. It is shown that the green line~(i.e., causal NEMTVII model at maximum compactness) satisfies all of the energy conditions while other causality-violating profiles at larger compactness only violate DEC.
\begin{figure}[h!]
	\centering
	\includegraphics[width=0.45\linewidth]{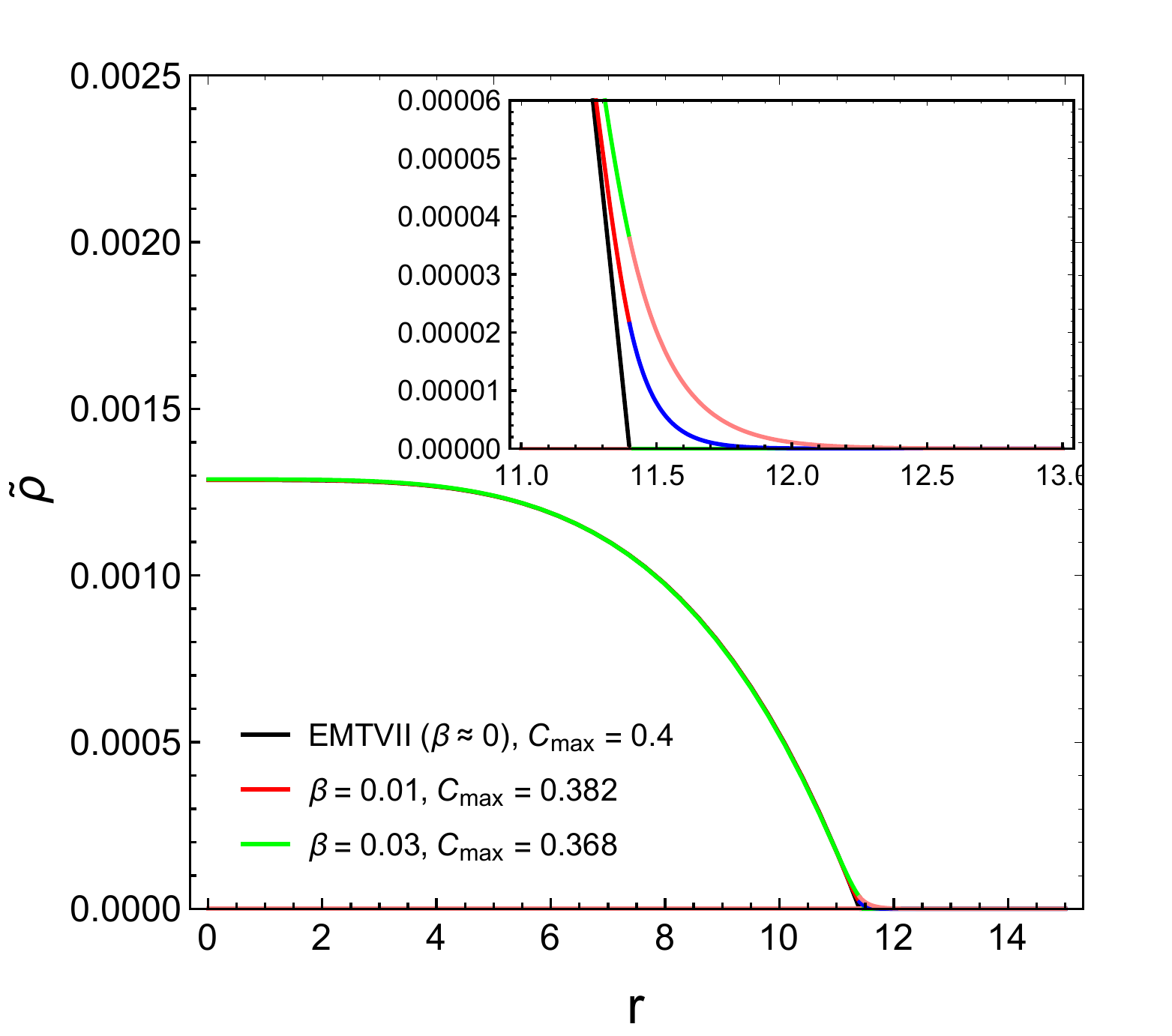}
	\includegraphics[width=0.45\linewidth]{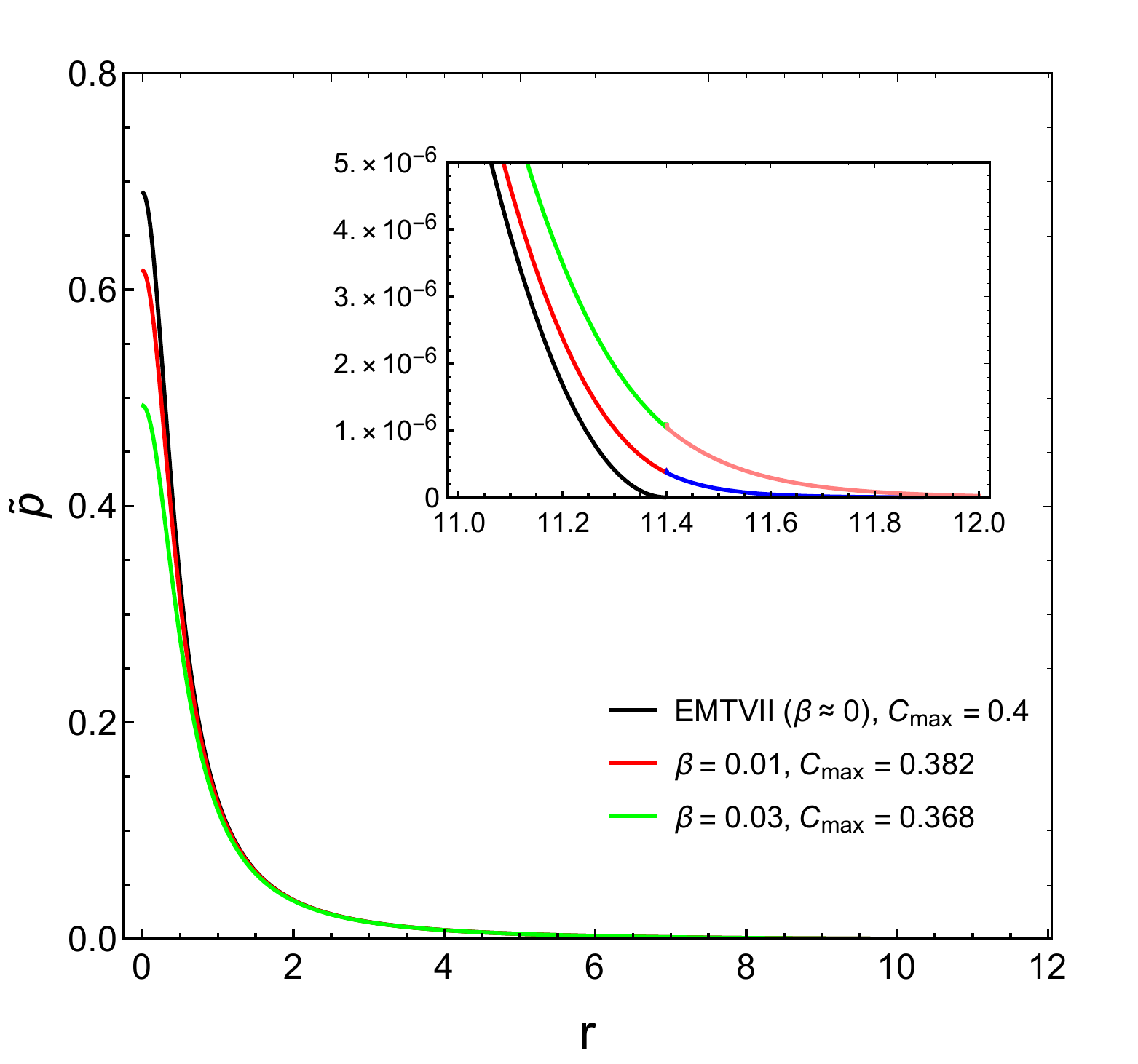}
	\caption{[Left] Nonlocal density profile with respect to radial coordinate $ r $. The red and green solid line represent the density with nonlocal gravity. Nonlocality notably smears out the surface of the star. $ \lbrace \beta,R_{N}/\textrm{Km}\rbrace =~\lbrace 0.01,~11.93 \rbrace,~\lbrace 0.01,~12.4 \rbrace,~\lbrace 10^{-12}~(\textrm{EMTVII}), 11.4  \rbrace$. [Right] The nonlocal pressure as a function of radius $ r $. $ \tilde{p}_{c} $ denotes the critical pressure and the surface is defined by the condition $ \tilde{p}(R_{N})=0. $}
	\label{fig:3}
\end{figure}

With the choices of parameter that we obtain in Fig.~\ref{fig:1}, we can proceed the analysis further by considering the profile of the nonlocal density as well as the pressure which are shown in Fig.~\ref{fig:3}. We also compare the nonlocal solution with the original EMTVII profile in order to see the effects of the nonlocal parameter. From Fig.~\ref{fig:3}, it is shown that the nonlocality allows the star to have a larger radius. The blue (red) and pink (green) lines are the $ \tilde{\rho}_{\textrm{ext}} $ ($ \tilde{\rho}_{\textrm{int}} $) profile. The more we incorporate the nonlocality $ \beta $, the larger the exterior part and thus a larger radius. The new radius of the smeared object, in the presence of $ \beta, $ is defined when the density approaches very small value, $\sim 10^{-7}$. For $ \beta=0.01, $ we have $ R_{N}=11.93~ \textrm{Km}. $ For $ \beta=0.03 $ we have $ R_{N}=12.4~ \textrm{Km} $ and no smeared surface appear in the EMTVII ($ \beta\approx 0 $) model. The nonlocal pressure with respect to radial coordinate is also plotted by solving the TOV equation in \eqref{11}.  However, if the compactness is larger than a certain maximum value~($\mathcal{C}>\mathcal{C}_{max} $), the function $ \nu(r) $ and $ \tilde{p}(r) $ become divergent. From Fig.~\ref{fig:3}, it can be inferred that nonlocality extends radius of the star.

\subsection{Mass-Radius~(MR) diagram in NEMTVII model}

\begin{figure}[h!]
	\centering
	\includegraphics[width=0.45\linewidth]{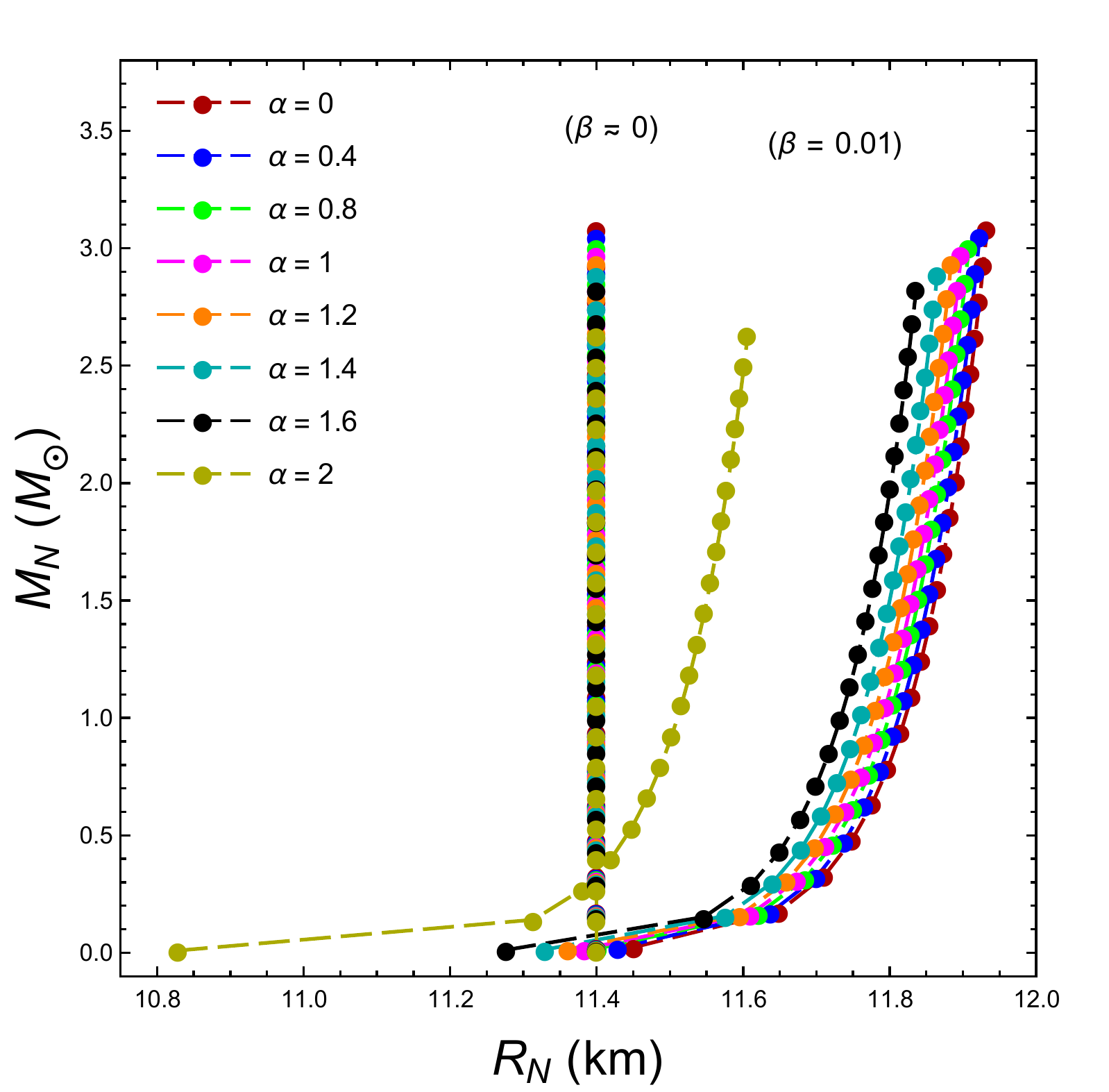}
	\includegraphics[width=0.45\linewidth]{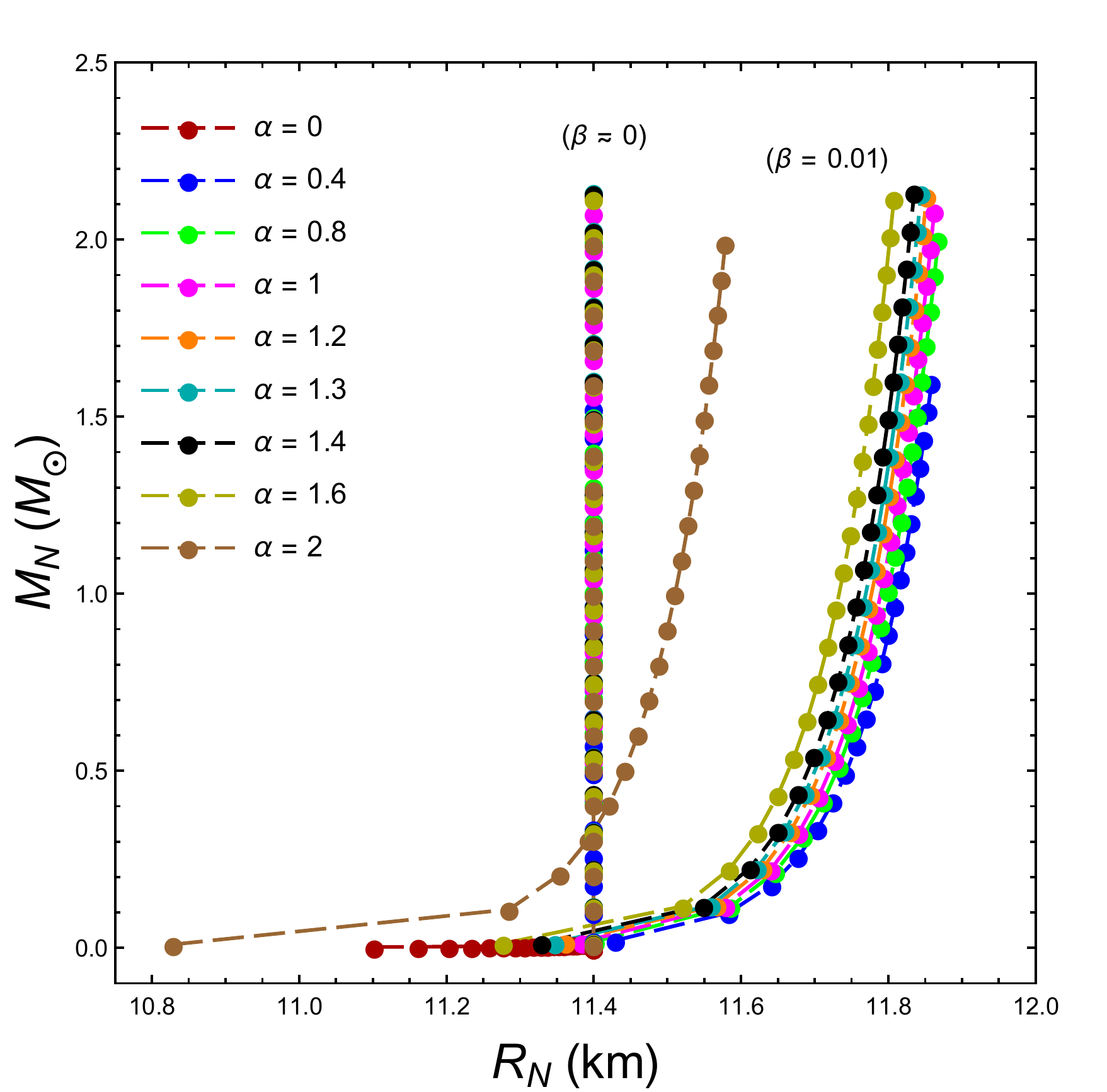}
	\caption{[Left] MR-diagram of NEMTVII star at maximum compactness, DEC and causality condition are not imposed. [Right] MR-diagram of NEMTVII star at maximum compactness under causality condition. Nonlocality smears the surface and enlarge the star radius for the massive branch. }
	\label{figMR}
\end{figure}

To understand the range of physical mass of star the NEMTVII mass model can provide, we present the MR diagram in Fig.~\ref{figMR}. In EMTVII model at fixed radius without the causality condition, increasing $\alpha$ reduces mass of the star and the maximum mass $M_{max}=3.1 M_{\odot}$ occurs at $\alpha = 0$. Under causality condition, however, the maximum mass $M_{max}\simeq 2.14 M_{\odot}$ occurs at $\alpha = 1.3-1.4$ and it is almost zero at $\alpha = 0$, see also Table~\ref{tab2}.   

When nonlocality is turned on, the star surface is smeared to larger radii as depicted in Fig.~\ref{figMR}. Interestingly, around $R\simeq 11.8$ km, NEMTVII mass model {\it satisfying causality condition} can cover stars with mass around $0.86-2.1 ~M_{\odot}$ for $\alpha = 0.4-1.6$.

\section{Gravitational Perturbation of the Ultracompact Star}  \label{SectIV}

One of the interesting aspect of ultracompact object with high compactness ($ 0.33\leq\mathcal{C}\leq 0.44 $) is the existence of light ring and the well of the effective potential inside the star \cite{CP2019}. The incoming waves can be trapped in the interior resulting in the gravitational waves in the quasinormal modes. We only focus on the axial gravitational perturbation considered in \cite{Chandrasekhar:1985kt,Chandrasekhar:1991fi}. The nonzero Ricci tensors from a perturbed metric are given in the following differential equations \cite{Chandrasekhar:1985kt,Chandrasekhar:1991fi}
\begin{eqnarray}
(e^{3\psi-\nu-\mu_{2}-\mu_{3}}Q_{23}),_{3} + e^{3\psi-\nu-\mu_{2}+\mu_{3}}Q_{02,0} &=& 0, ~~~~~~~~~~~~~ ~~~~~~~~~~~~~~~~\label{pereq1} \\ (e^{3\psi-\nu-\mu_{2}-\mu_{3}}Q_{23}),_{2} - e^{3\psi-\nu+\mu_{2}-\mu_{3}}Q_{03,0} &=& 0,~~~~~~~~~~~~~~~~ ~~~~~~~~~~~~~~~~\label{pereq2}
\end{eqnarray}
where $ Q_{AB}= q_{A,B} - q_{B,A} $ and $ Q_{A0} = q_{A,0}-\omega_{,A}~~(\textrm{A}=2,3). $ The variable $ \nu,~\mu_{2},~\mu_{3},~\textrm{and } \psi$ are the unperturbed metric functions and $\omega, q_{2}, q_{3}$ are the perturbations defined in the following metric~\cite{Chandrasekhar:1985kt,Chandrasekhar:1991fi},
\begin{equation}
ds^{2} = e^{2\nu} dt^{2}-e^{2\psi}(d\phi-\omega~ dt-q_{2}dx^{2}-q_{3}dx^{3})^{2}-e^{2\mu_{2}}(dx^{2})^{2}-e^{2\mu_{3}}(dx^{3})^{2},
\end{equation}
where $x^{2}, x^{3}= r, \theta$ respectively. From the integrability of (\ref{pereq1}) and (\ref{pereq2}), we obtain
\begin{equation}
\label{req}
[e^{-3\psi+\nu+\mu_{2}-\mu_{3}}(e^{3\psi+\nu-\mu_{2}-\mu_{3}}Q_{23}),_{3}],_{3} + [e^{-3\psi+\nu+\mu_{2}+\mu_{3}}(e^{3\psi+\nu-\mu_{2}-\mu_{3}}Q_{23}),_{2}],_{2} = Q_{23,0,0}.
\end{equation}
Eqn.~\eqref{req} can be separated by expanding in terms of Gegenbauer function $ \mathcal{C}_{n}^{\alpha}(\theta), $ which obeys
\begin{equation}
\label{eq}
\left[\frac{d}{d\theta}\sin^{2\alpha}\theta\frac{d}{d\theta}+n(n+2\alpha)\sin^{2\alpha}\theta\right]\mathcal{C}_{n}^{\alpha}(\theta)=0.
\end{equation}
By separation of variables
\begin{eqnarray}
e^{3\psi+\nu-\mu_{2}-\mu_{3}}Q_{23}=r~ \Psi(r)~ \mathcal{C}_{l+2}^{-3/2}(\theta)~~~~\textrm{and introducing}~~~~\frac{dr_{*}}{dr} = e^{-\nu+\mu_{2}},
\end{eqnarray}
the tortoise coordinate, Eqn.~\eqref{req} can be reduced to the Schrodinger-like wave equation in $ r_{*} $ as
\begin{equation}
\label{pdp}
\left[\frac{\partial^2}{\partial t^2}-\frac{\partial^2}{\partial r_{*}^2}+V(r)\right]\Psi(r_{*},t)=0,
\end{equation}
where the effective potential is 
\begin{equation}
\label{veff}
V(r) = \frac{e^{\nu}}{r^3}\left[l(l+1)r+ 4\pi r^3\left(\tilde{\rho}-\tilde{p}\right)-6\tilde{m}(r)\right].
\end{equation}
This master equation is often called time-dependent Regge-Wheeler (RW) equation which can be analyzed in terms of the tortoise coordinate ($ r_{*} $)
\begin{equation}
r_{*}= \int_{0}^{r} \sqrt{-\frac{g_{rr}}{g_{tt}}} ~dr,
\label{tortoise}
\end{equation}
where $ g_{tt} $ and $ g_{rr} $ are the metrics of spacetime both inside and outside of the star. 
\begin{figure}[h!]
	\centering
	\includegraphics[width=0.6\linewidth]{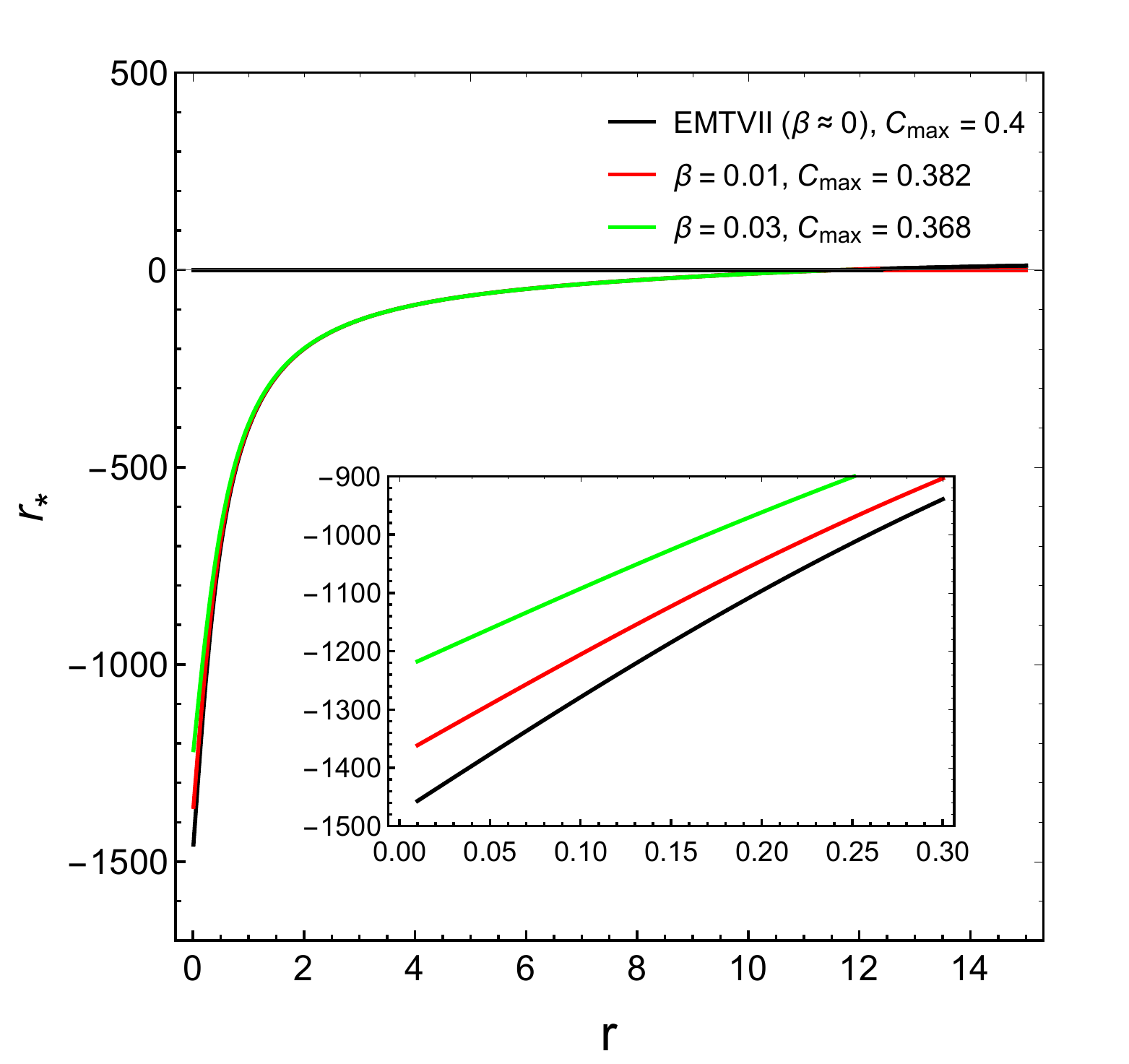}
	\caption{Tortoise coordinate vs radial coordinate}
	\label{fig:4}
\end{figure}
Fig.~\ref{fig:4} represents the relation between $ r_{*} $ and $ r $ for the cases $ \mathcal{C}_{max} = 0.368  $ with $ \beta=0.03 $, $ \mathcal{C}_{max} = 0.382  $ with $ \beta = 0.01 $, and EMTVII parameter. The value of $ r_{*}(0) $ is finite and becomes more negative as compactness increases. From this aspect, it can be inferred that compactness plays an important role on the ``tortoise-width" of the interior of the star. Furthermore, the exterior region of the star exhibits linear relation between $r_{*}$ and $r$.
\begin{figure}[h!]
	\centering
	\includegraphics[width=0.496\linewidth]{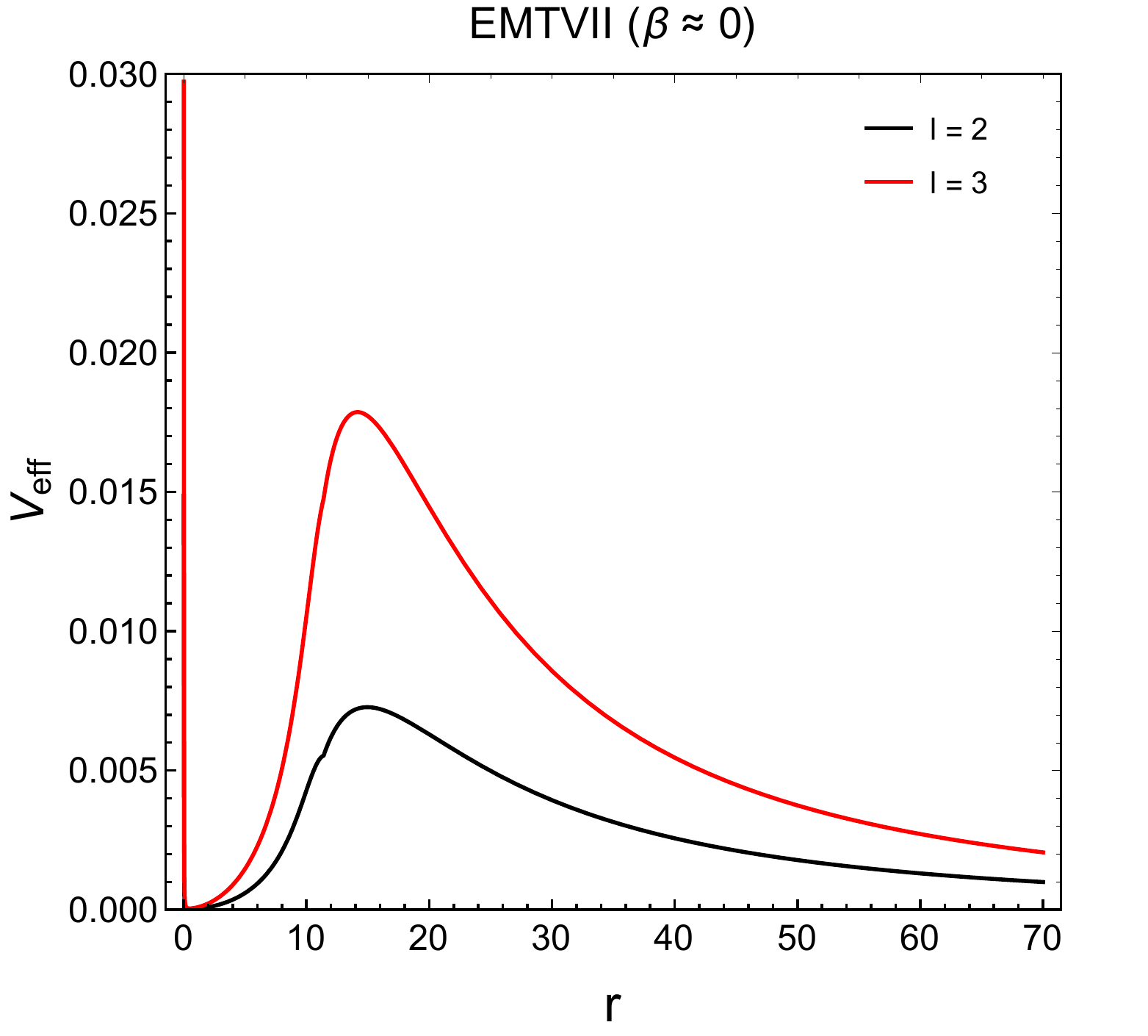}
	\includegraphics[width=0.496\linewidth]{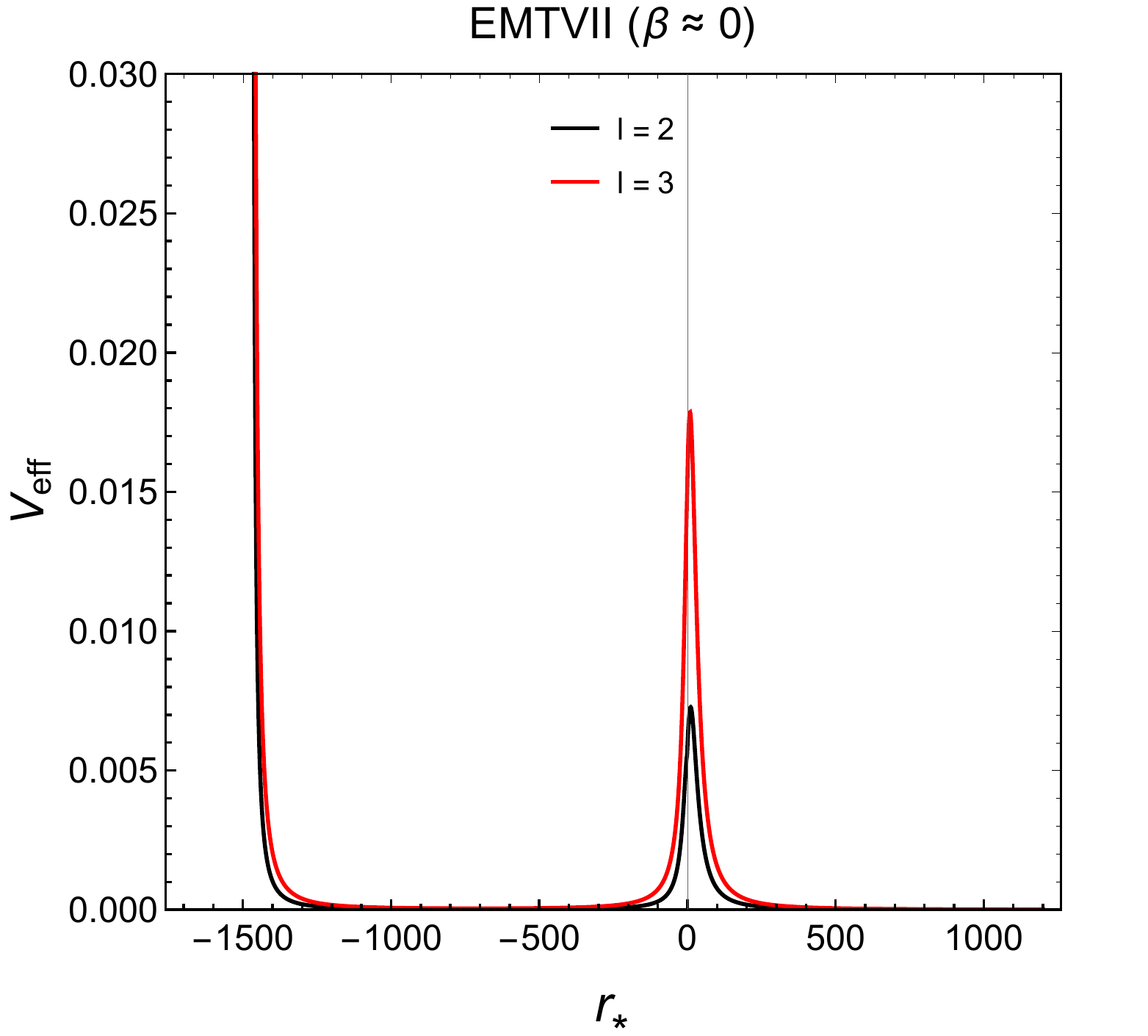}
	\caption{Effective potential ($ V_{eff} $) versus radial coordinate, $ r, $ and the tortoise coordinate, $ r_{*} $. The black and red solid line represents the EMTVII for $ l=2 $ and $ l=3 $ respectively.}
	\label{fig:5}
\end{figure}

In Fig~\ref{fig:5}, we show the effective potential as the function of the radius, $ r, $ as well as the tortoise coordinate, $ r_{*}. $ Here, we consider first the EMTVII model in the compactness range of ultracompact objects i.e., $ 0.33\leq \mathcal{C} \leq 0.44 $. The star surface is located at $ R <3M $, and the exterior solution has a barrier at $ R \approx 3M. $ A perturbed star's effective potential also has the infinite potential value at the center due to the centrifugal contribution in the first term on the right-hand side~(RHS) of Eqn.~\eqref{veff}. In the tortoise coordinate, the light ring occurs almost at the peak of the potential. The effective potential also possess a second light ring at the minimum. This light ring should be stable since $ V_{eff}''>0. $ Moreover, $ l $ in the potential determines the height of the barrier. Later, we will see that the quasinormal mode also depends on $ l $.
\begin{figure}[h!]
	\centering
	\includegraphics[width=0.497\linewidth]{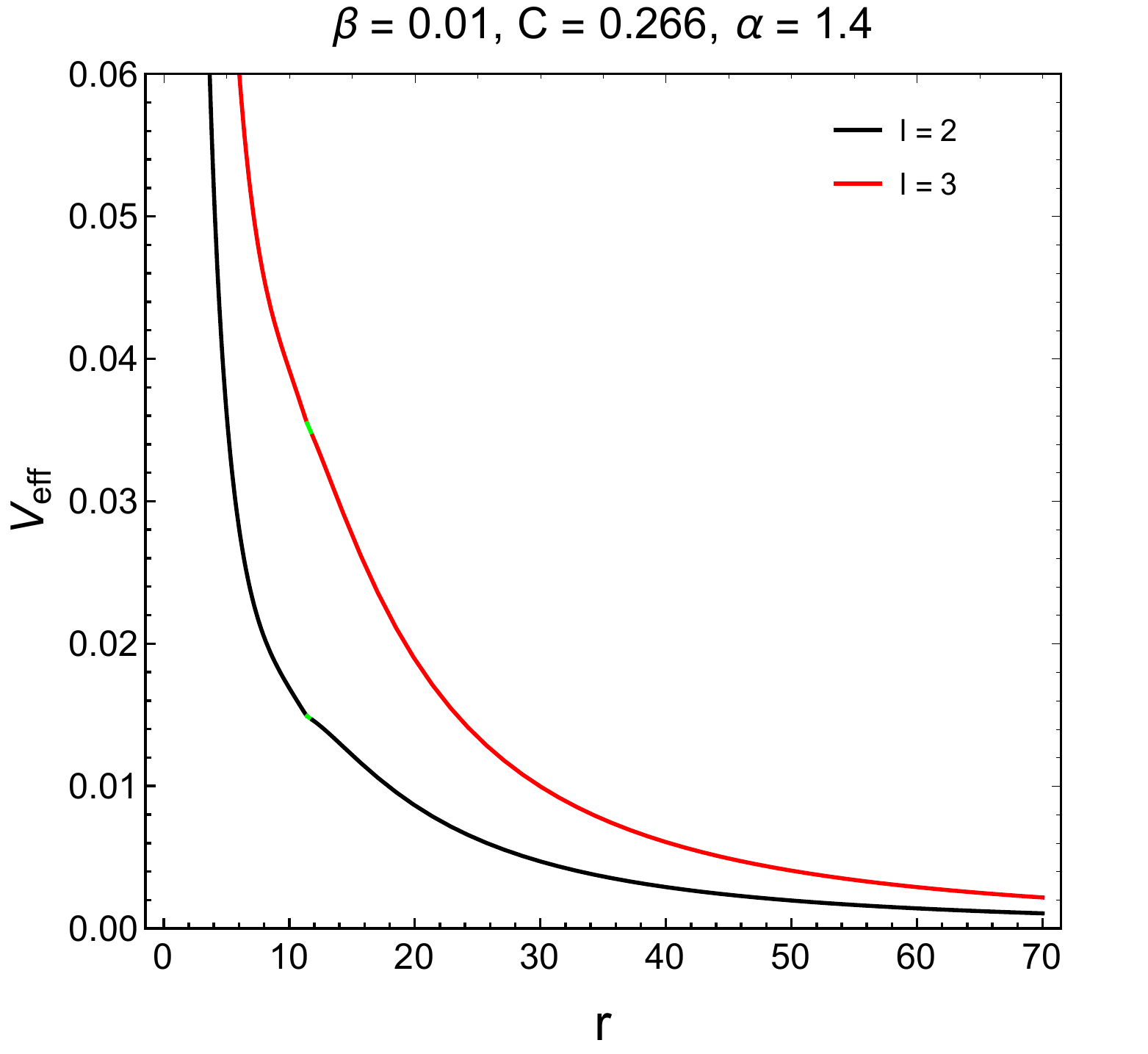}
	\includegraphics[width=0.496\linewidth]{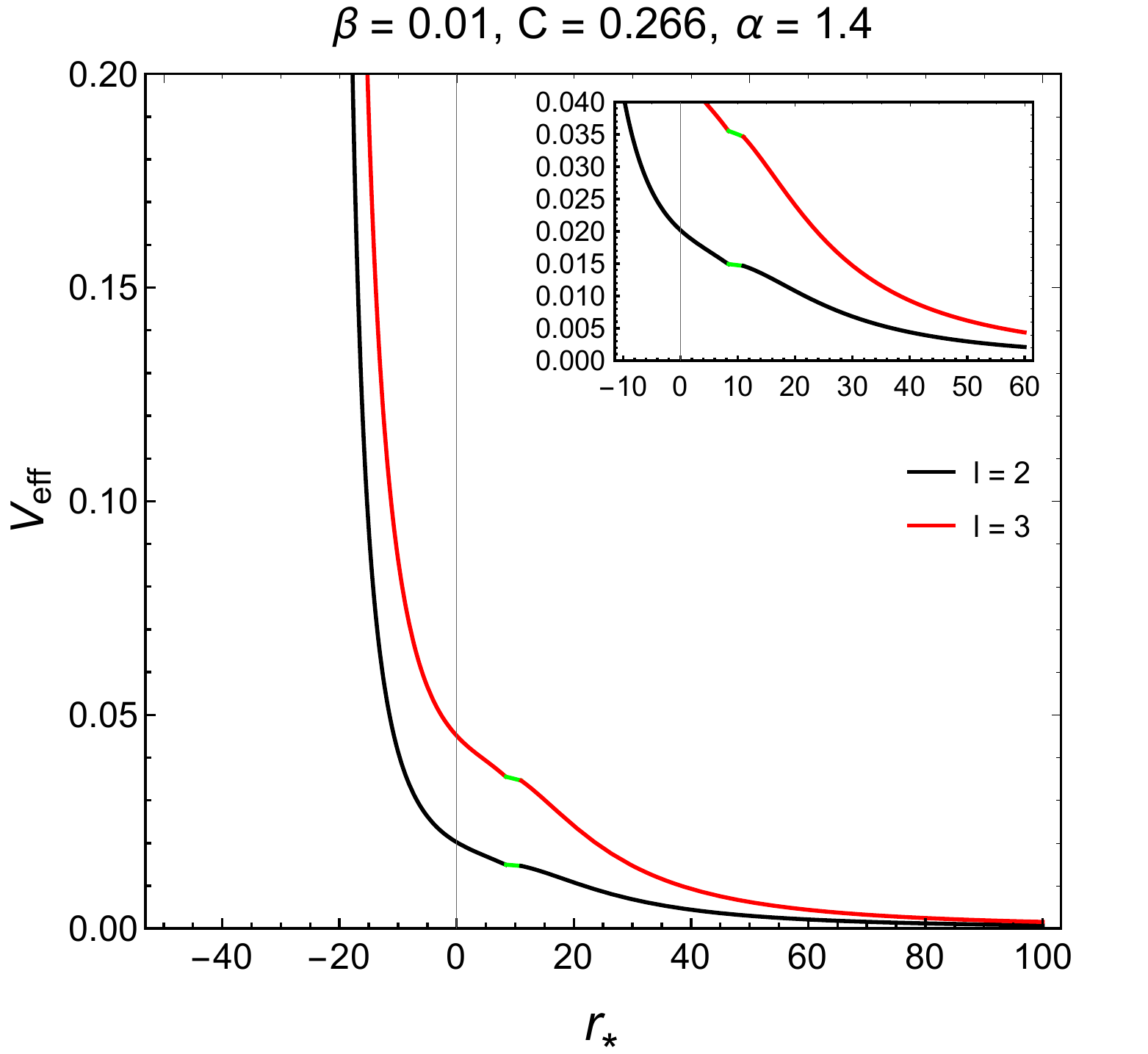}
	\caption{Effective potential ($ V_{eff} $) with respect to radial $ r, $ and the tortoise coordinate, $ r_{*} $ saturating the causal limit. The figures represent the profile within the causal limit for $ l=2 $ and $ l=3 $.}
	\label{fig:6}
\end{figure}
\begin{figure}[h!]
	\centering
	\includegraphics[width=0.497\linewidth]{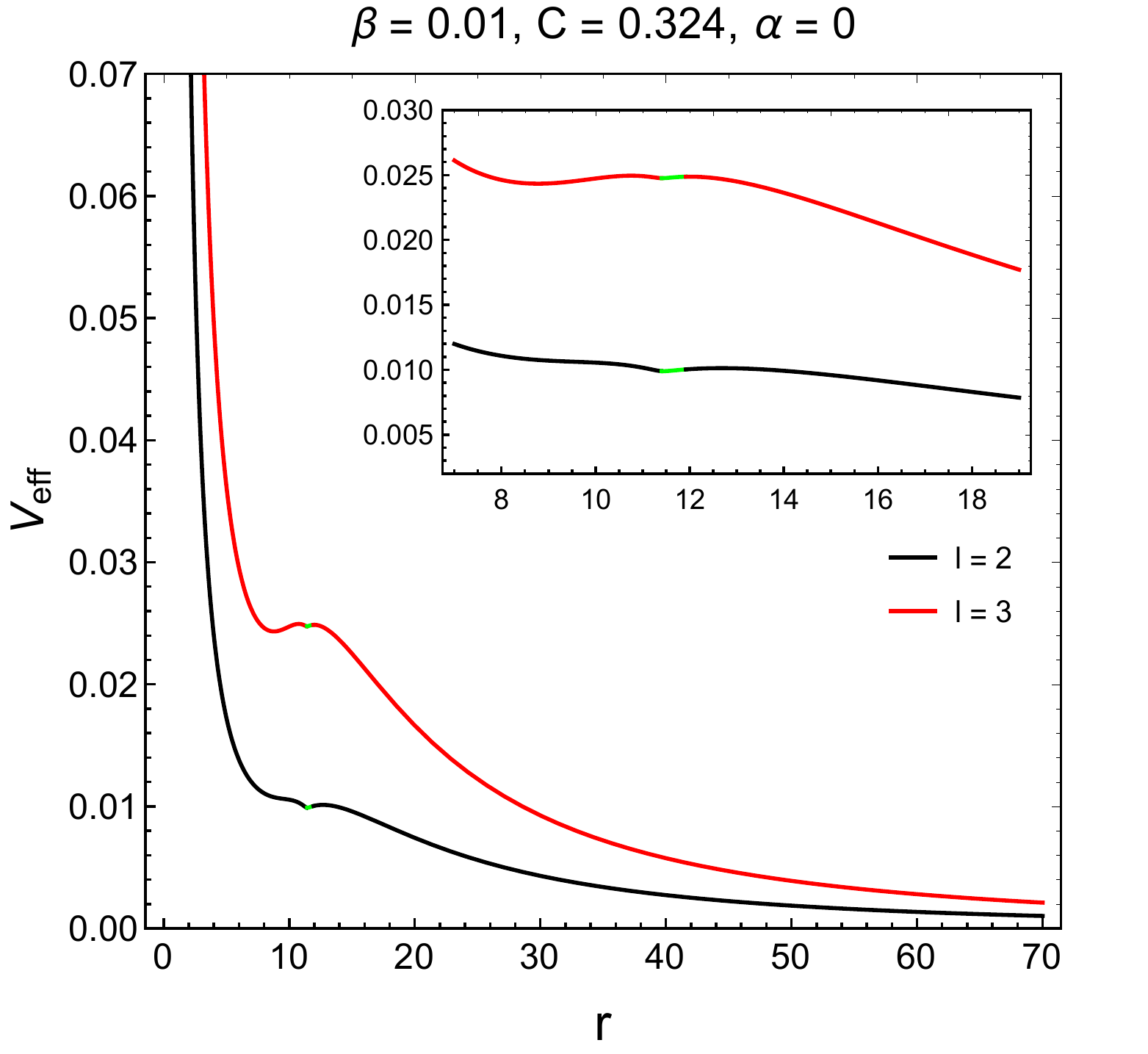}
	\includegraphics[width=0.496\linewidth]{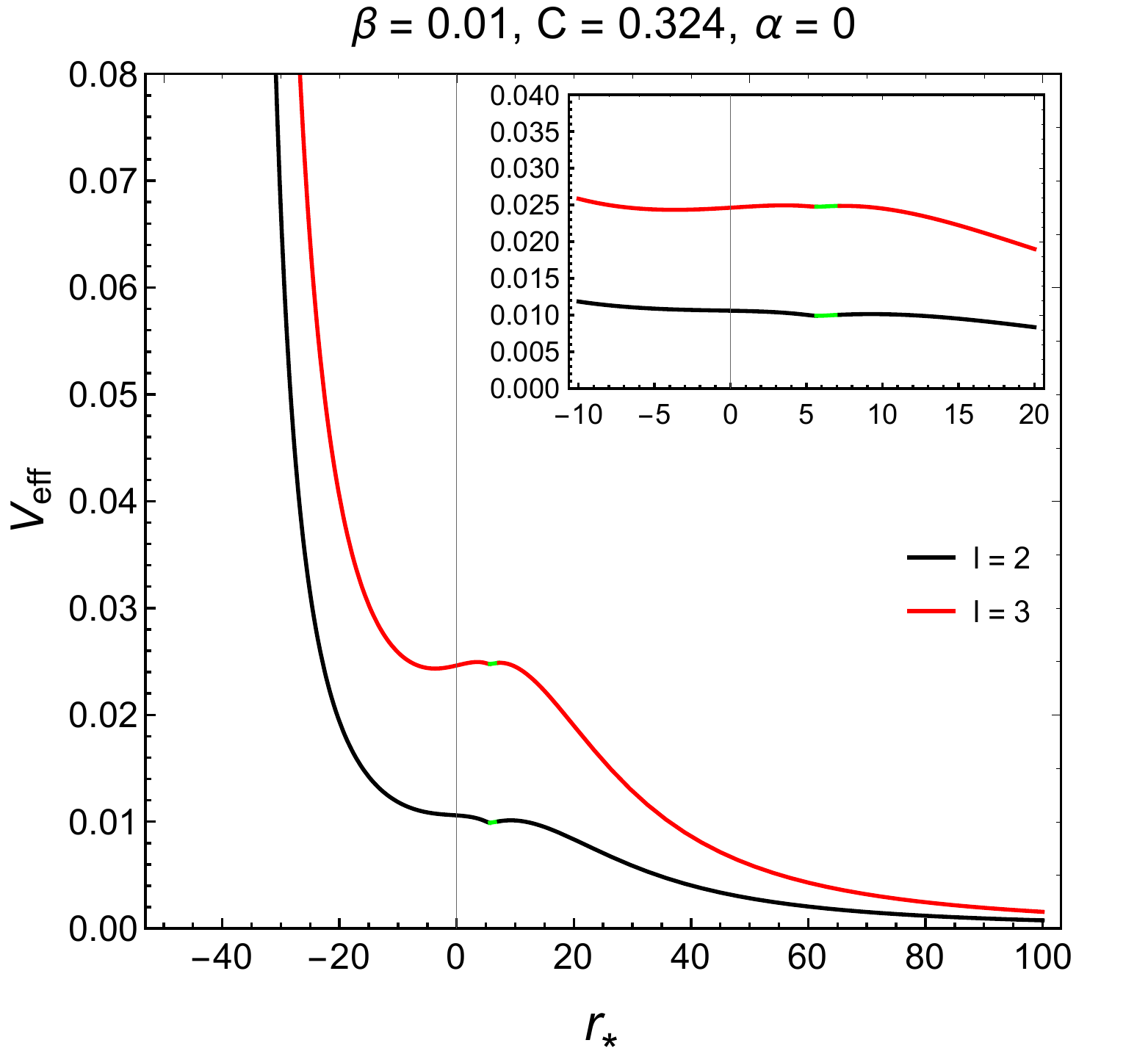}
	\caption{Effective potential ($ V_{eff} $) with respect to radial $ r, $ and the tortoise coordinate, $ r_{*} $ saturating the DEC. }
	\label{fig:6.5}
\end{figure}

On the other hand, Fig.~\ref{fig:6} shows effective potential in the presence of nonlocal gravity at $\mathcal{C}_{max}$ with causality condition imposed~(see Right plot in Fig.~\ref{fig:1}). The upper (lower) part represents the nonlocal potential with $ \beta=0.01,~\alpha=1.4,~\mathcal{C}=0.2667$ for $l=3~(2)$ respectively. Since the compactness is below the ultracompact region, the light ring does not exist for both $ l. $ The potential well rapidly disappears and does not have sufficient width and depth to produce quasinormal modes and gravitational echoes. Similarly for DEC saturating star in NEMTVII model shown in Fig.~\ref{fig:6.5}, the effective potential well is too shallow to sustain echoes. 

\begin{figure}[h!]
	\centering
	\includegraphics[width=0.497\linewidth]{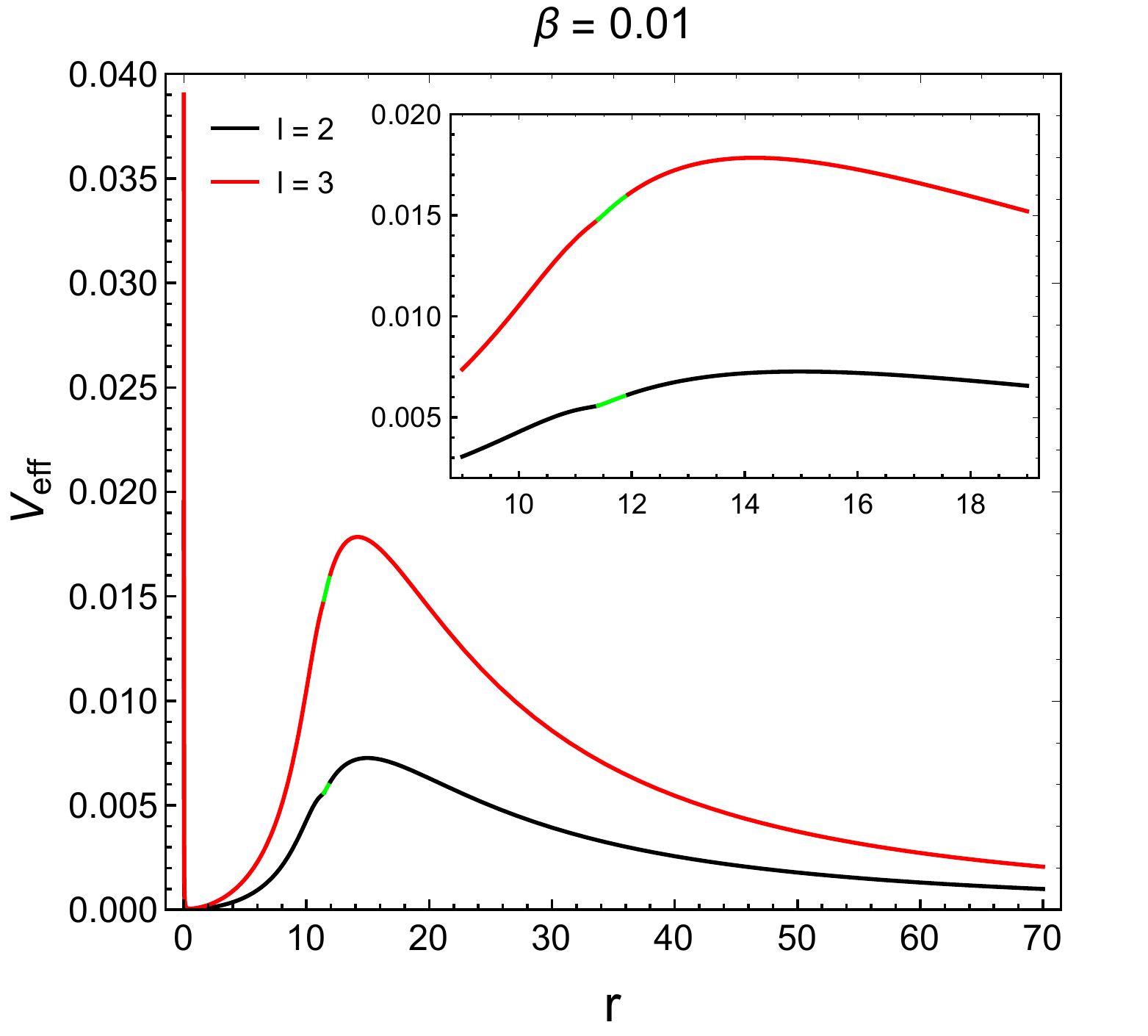}
	\includegraphics[width=0.496\linewidth]{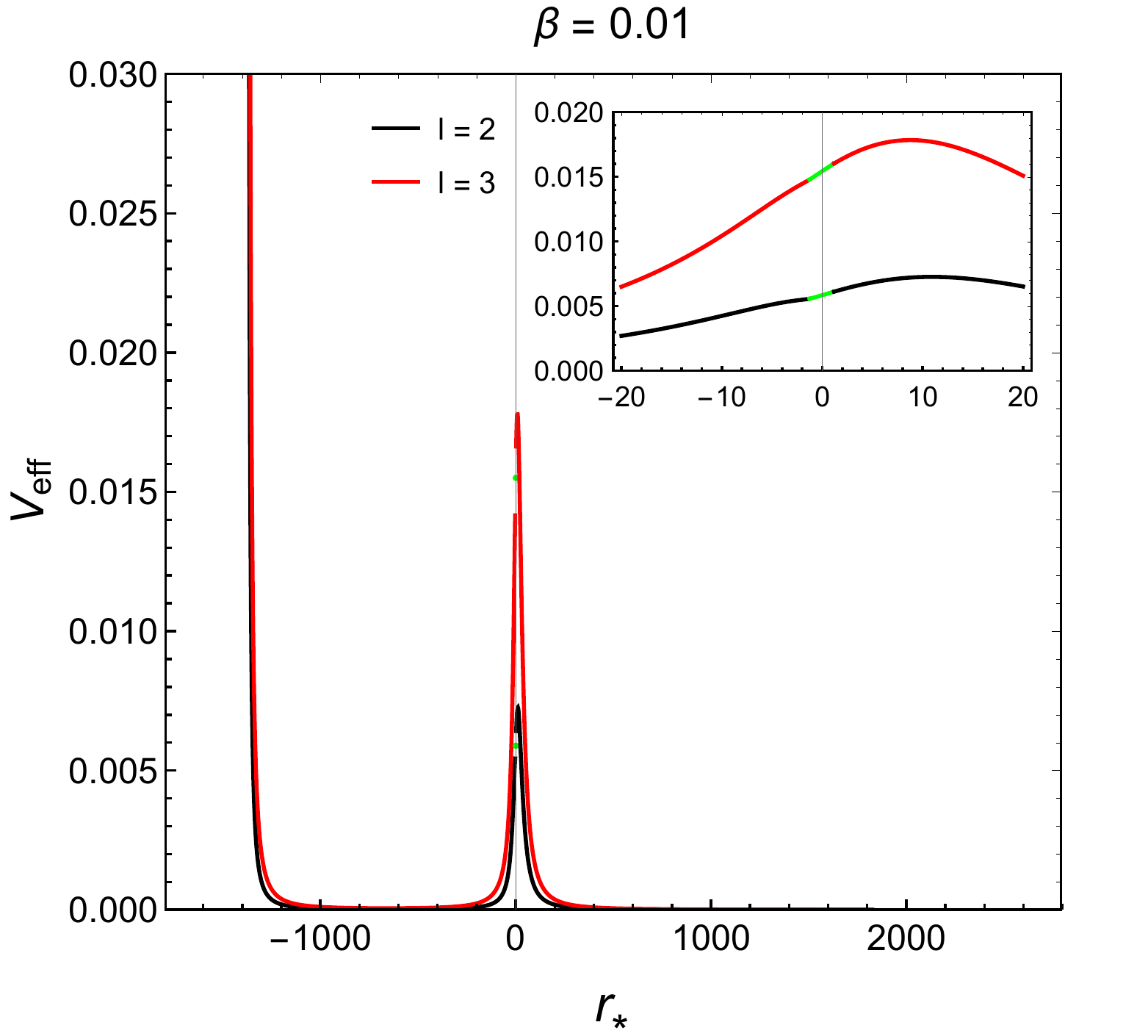}
	\includegraphics[width=0.497\linewidth]{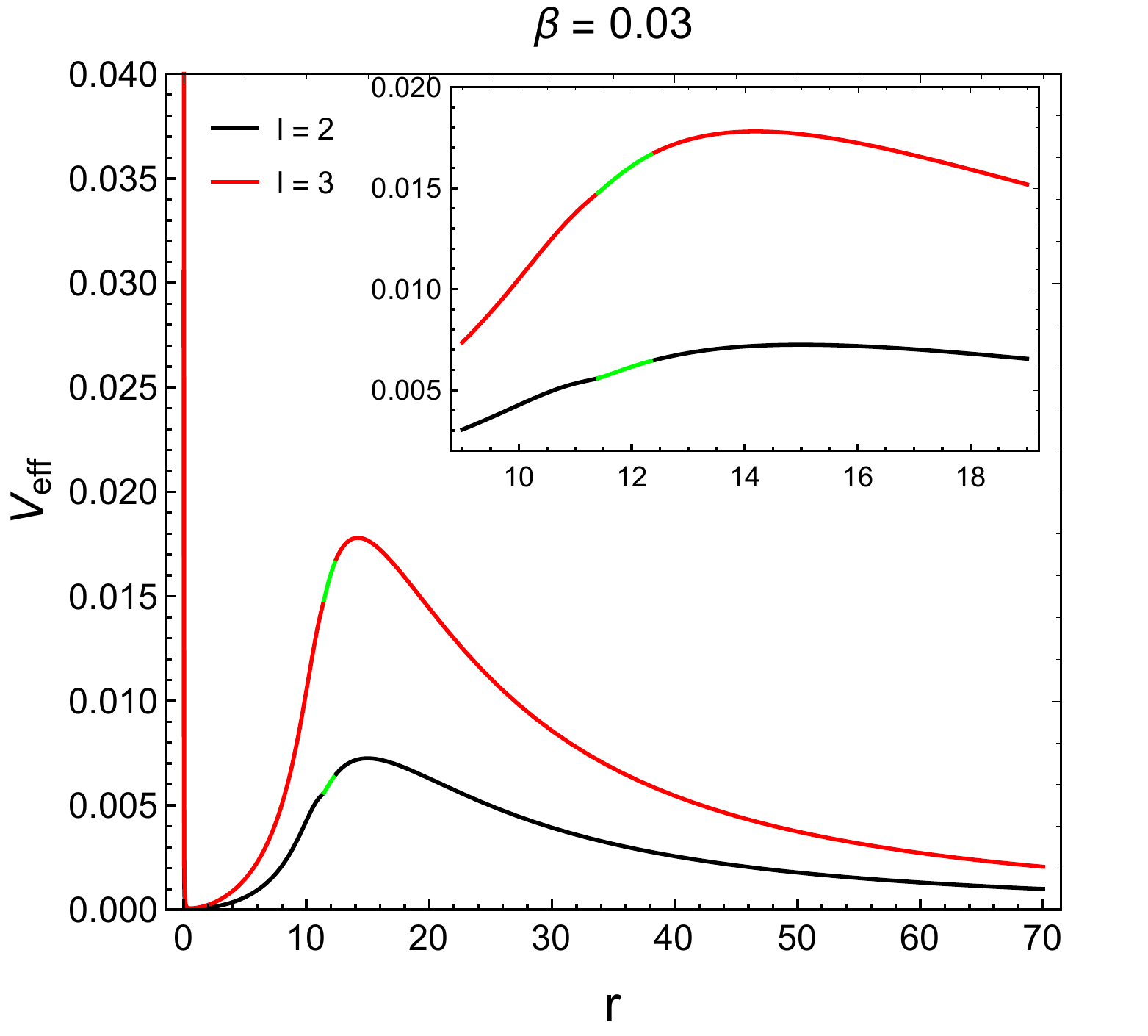}
	\includegraphics[width=0.496\linewidth]{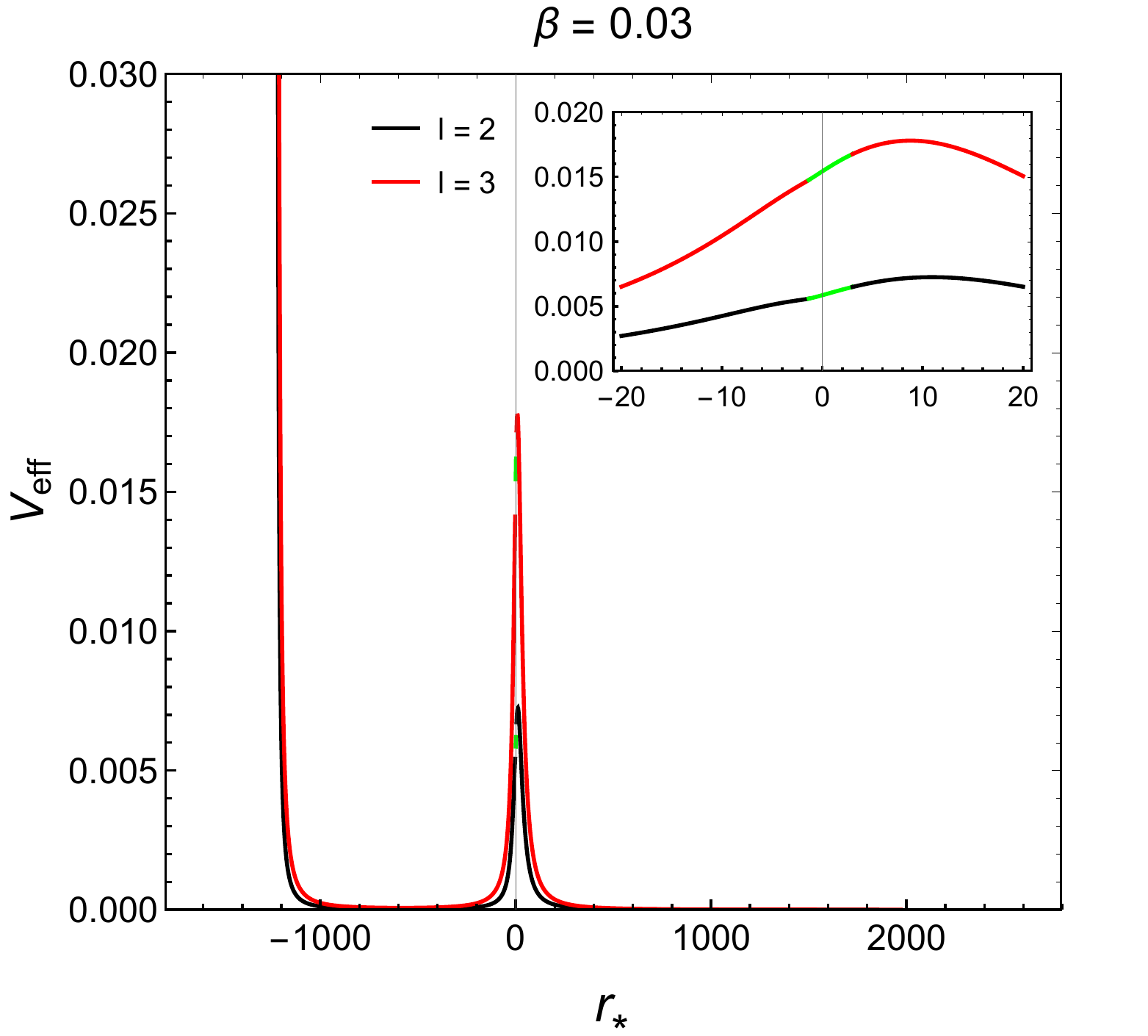}
	\caption{Effective potential ($ V_{eff} $) versus radial coordinate $ r, $ and the tortoise coordinate, $ r_{*} $. The upper (lower) curve represents the $ V_{eff} $ profile with $ \beta=0.01,~\alpha=0,~\mathcal{C}_{max}=0.382 $ ($\beta=0.03,~\alpha=0,~\mathcal{C}_{max}=0.368 $) for $l=3~(2)$ respectively. }
	\label{fig:7}
\end{figure}

However, if causality condition is not imposed, generically the nonlocal EoS of NEMTVII with $ \mathcal{C}_{max} $ provides a well potential with surface at $ R_{N}<3M $ as presented in Fig.~\ref{fig:7}. Notably, the larger the nonlocal parameter, the wider the smeared surface (green colour) becomes. The whole interior of the star is within $ 0\leq r \leq R_{N} $. For $ r>R_{N} $ we match it with the exterior Schwarzschild $ V_{\textrm{eff}} $ profile. The existence of potential well is crucial for the production of gravitational echoes. The deeper the well, the more pronounced echoes could be produced. The wider the well, the longer it will take for the echo to propagate through the interior of the ultracompact star.

\subsection{Quasinormal Modes} \label{SectQNM}

The literatures from the early stage \cite{Kojima:1995nc,Andersson:1995ez} categorize the QNM spectrum related to the effective potential into three kinds of spacetime modes, i.e., the trapped modes~\cite{Kokkotas:1994an} produced by potential well of ultracompact stars, $w$-modes~\cite{Kokkotas:1992xak} produced from scattering on the top of the potential barrier, and $w_{II}$ mode~\cite{Leins:1993zz} that corresponds to the extremely fast damped mode. In this section, we will discuss only the crucial mode for ultracompact stars (axial part), i.e., the trapped modes.

For the waves with definite energy and possible damping,~$ \Psi(r_{*},t)=\psi(r)e^{-i\omega_{n} t} $, Eqn. (\ref{pdp}) can be cast into
\begin{equation}
\label{TI}	
\frac{d^2 \psi}{dr_{*}^2} + [\omega^2_{n}-V_{\textrm{eff}}(r)]\psi=0,
\end{equation}
where $ \omega_{n} $ is the QNM and $ \psi(r) $ is the eigen solution of the time independent RW equation. QNMs, in general, encode dissipative property of spacetime and play a crucial role in the formation of gravitational wave echoes when stars reach the final stage during ringdown phase. When the star are in these states, the unstable circular orbit traps the primary signal. In order to obtain the QNMs, the complex eigenvalues on the time-independent equation shown in \eqref{TI} must be solved with the appropriate boundary conditions. From the black hole point of view \cite{Urbano:2018nrs}, since nothing can escape from it, we must employ first, an incoming spherical boundary condition at the black hole horizon and second, an outgoing waves at spatial infinity \cite{Urbano:2018nrs}. However, when we consider the ultracompact or compact stars the boundary conditions will be different since they lack event horizon. The first boundary condition should be replaced by the regularity condition at the center, whereas the second boundary condition is unaltered. 

Here, we employ Volkel and Kokkotas's \cite{Volkel:2017ofl} procedure to obtain the QNMs using Wentzel-Kramers-Brillouin (WKB) approximation. In quantum mechanics, Bohr-Sommerfeld (BS) rule is a well-known method to approximate the energy spectrum, $ E_{n}, $ of bound states in a potential. With the WKB theory, it is possible to include higher-order correction of the BS rule \cite{popov} as follows~\cite{atomicwkb}
\begin{equation}
\label{BS}
\int_{x_{0}}^{x_{1}} \sqrt{E_{n}-V(x)} dx = \pi \left(n + \frac{1}{2}\right) - \frac{i}{4} e^{2i\int_{x_{1}}^{x_{2}} \sqrt{E_{n}-V(x)} ~dx},
\end{equation}
where $ x_{0} $ and $ x_{1} $ are the classical turning point(s) determined by the roots of the integrand and also depend on the energy spectrum. The second term in Eqn. \eqref{BS}  is the additional term in the generalized BS rule, where $ x_{2} $ is the third classical turning point on the right side of the potential barrier\footnote{For a detailed discussion of this procedure, see Ref.~\cite{Volkel:2017ofl}.}. This additional term denotes an exponentially small imaginary part of the energy spectrum, which measures the barrier penetrability. We can simplify further by writing the energy spectrum into $ E_{n}=E_{0n}+iE_{1n} $, where $ E_{0n} $ is the real part of the energy whereas $ E_{1n} $ is small imaginary energy. Insert these energy spectra to the left-hand side of Eqn.~\eqref{BS} and match with the real and imaginary parts on the right-hand side of Eqn.~\eqref{BS}. The results can be written as
\begin{equation}
\label{e0n}
\int_{x_{0}(E_{0n})}^{x_{1}(E_{0n})} \sqrt{E_{0n}-V(x)} dx = \pi \left(n + \frac{1}{2}\right),
\end{equation}
and
\begin{equation}
\label{e1n}
E_{1n} = -\frac{1}{2}\left( \int_{x_{0}(E_{0n})}^{x_{1}(E_{0n})} \frac{1}{\sqrt{E_{0n}-V(x)}} dx \right)^{-1} e^{2i \int_{x_{1}(E_{0n})}^{x_{2}(E_{0n})} \sqrt{E_{0n}-V(x)} ~ dx}.
\end{equation}
Furthermore, we use the well known analytic fitting potential described below \cite{Volkel:2017ofl}
\begin{eqnarray}
U_{Q} = U_{0}+\lambda_{0}^2 (x-x_{min})^{4}~~~ \textrm{and}~~~ U_{BW} = \frac{U_{1}}{1+\lambda_{1}(x-x_{max})^2}.
\end{eqnarray} 
Both functions above are called quartic oscillator potential and Breit-Wigner potential respectively. These two functions depend on $ U_{0}, $ $ U_{1}, $ $ \lambda_{0}, $ and $ \lambda_{1}$ parameters. The value of these parameters is then matched with the true effective potential (in tortoise coordinate). $ U_{0} $ is the minimum of the true potential, whereas $ U_{1} $ is the maximum of the true potential. Next, $ \lambda_{1} $ can be obtained by identifying $ V_{max}'' $ on the $ U_{BW} $ function, which can be written as $ \lambda_{1}=-V_{max}''/2V_{max}. $ The last parameter, $ \lambda_{0}, $ we demand that the quartic oscillator must be equal to the Regge-Wheeler equation at the surface \cite{Volkel:2017ofl}. We can evaluate the energy spectrum by substituting both fitting functions into Eqn.\eqref{e0n} and Eqn.\eqref{e1n}. The detailed calculation of the integral can be found in Ref.~\cite{Volkel:2017ofl}. The final solution reads
\begin{eqnarray}
\label{E0n}
E_{0n} &=& U_{0} + \lambda_{0}^{2/3} \left[\frac{3\pi}{4~ \mathcal{K}(-1)} \left(n+\frac{1}{2}\right)\right]^{4/3}, \\ E_{1n} &=& -\frac{\sqrt{\lambda_{0}}(E_{0n}-U_{0})^{1/4}}{4~ \mathcal{K}(-1)} ~\textrm{Exp}\left[4i  \sqrt{\frac{U_{1}-E_{0n}}{\lambda_{1}}}~ \mathcal{E}\left( i \sinh^{-1}\left(\sqrt{\frac{U_{1}}{E_{0n}}-1}\right),\frac{E_{0n}}{E_{0n}-U_{1}} \right)   \right],\nonumber \\
\label{E1n}
\end{eqnarray}
where $ \mathcal{K}(a) $ is the complete elliptic integral of the first kind and $ \mathcal{E}(a,b) $ denotes the elliptic integral of the second kind \cite{Abraham}. The variable $ n $ shown in the Eqn. \eqref{E0n} is the overtone number. The numerical results from Eqn. \eqref{E0n} and Eqn. \eqref{E1n} are the same to those obtained from the formula given in Ref.~\cite{Volkel:2017ofl}.
\begin{figure}
	\centering
	\includegraphics[width=0.6\linewidth]{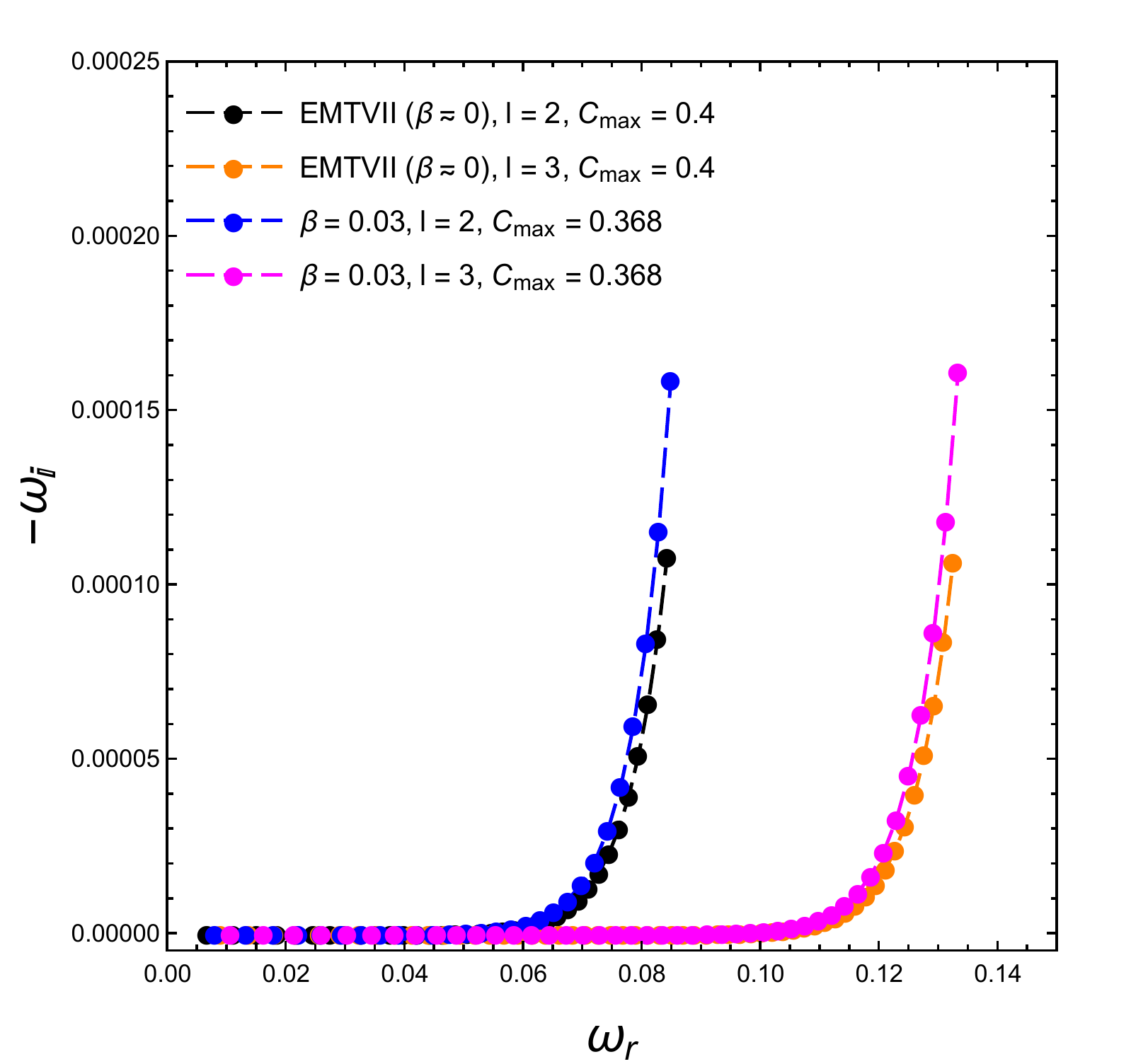}
	\caption{Quasinormal modes $ |\omega_{i}| $ vs $ \omega_{r} $. The exact values are presented in \ref{AppC}.}
	\label{fig:8}
\end{figure}

We can match the energy with the complex eigenvalue shown in Eqn.\eqref{TI} which can be written as, $ \omega_{n}=\sqrt{E_{n}}=\sqrt{E_{0n}+iE_{1n}}=\omega_{r}+i\omega_{i}$, where the real part describes the frequency of the oscillation and the imaginary part describes the inverse of the damping time, $ \tau_{d} $. This brings us to the approximate analytic results for all trapped modes in ultracompact stars.

The result of the QNMs is shown in Fig.~\ref{fig:8}. We plot the QNM, $ |\omega_{i}| $ vs $ \omega_{r} $, with different $ l $ and maximum compactness. The number of the dots in the plot represents the overtone number. For the original EMTVII, the overtone number for $ l=2 $ exceed $ n\sim 35 $ and for $ l=3 $ we have $ n \sim 55$. For NEMTVII ($ \beta = 0.03 $), the overtone number for the lowest mode is $ n\sim 27 $ and for $ l=3 $ is $ n\sim 43 $. From these overtone numbers, we can conclude that both $ l=2 $ and $ l=3 $ modes allow the ultracompact star EMTVII to have more trapped modes than the NEMTVII counterpart. On the other hand, for both $ l $, the imaginary parts increase as the compactness decreases (from EMTVII to NEMTVII). The horizonless ultracompact object with EMTVII matter allows the axial perturbations to persist with damping time longer than the NEMTVII model. This is the consequence of the fact that the well of the EMTVII is wider than the NEMTVII.

\subsection{Gravitational Echoes and Echo Time}

\begin{figure}[h!]
	\centering
	\includegraphics[width=0.45\linewidth]{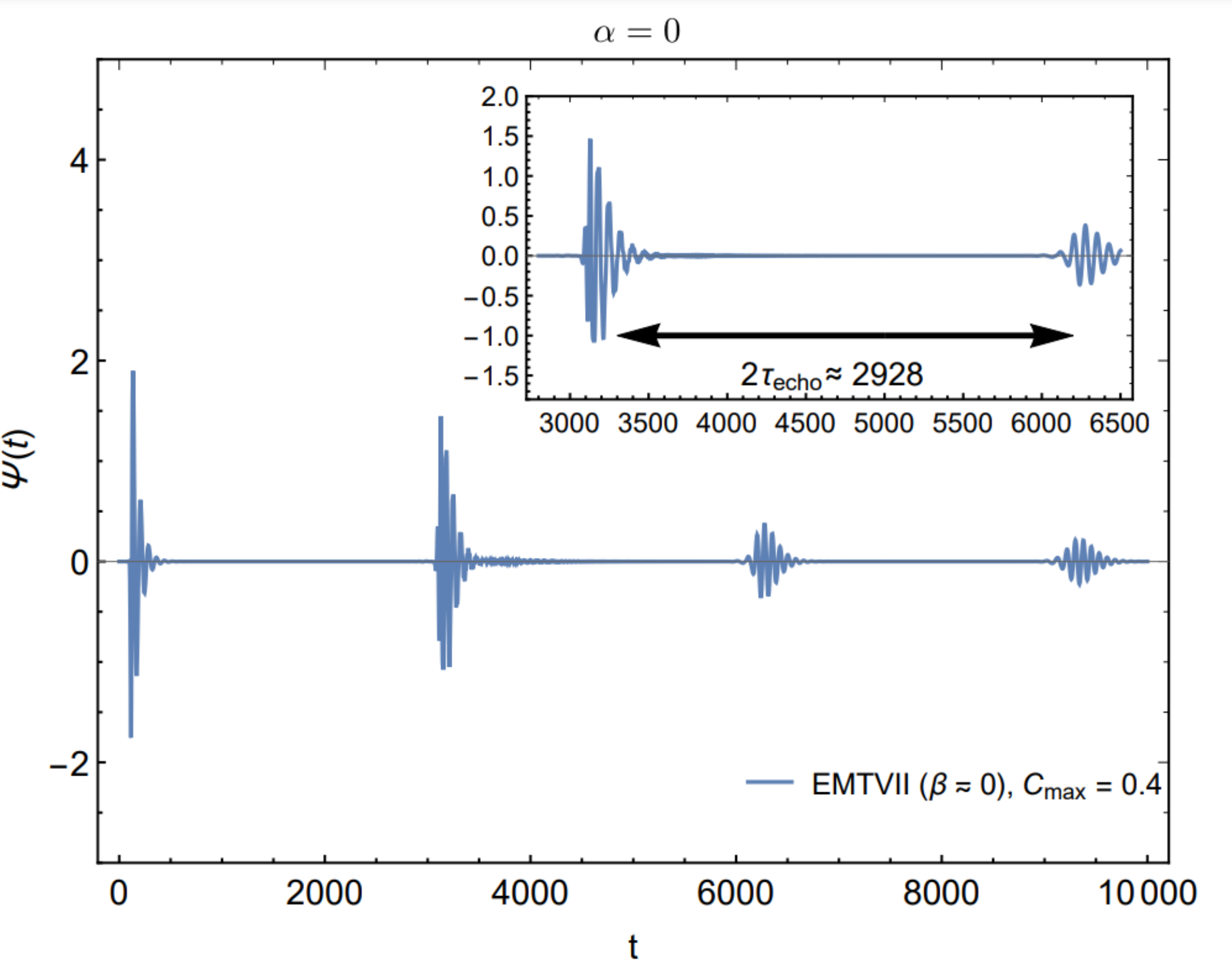}
       \includegraphics[width=0.45\linewidth]{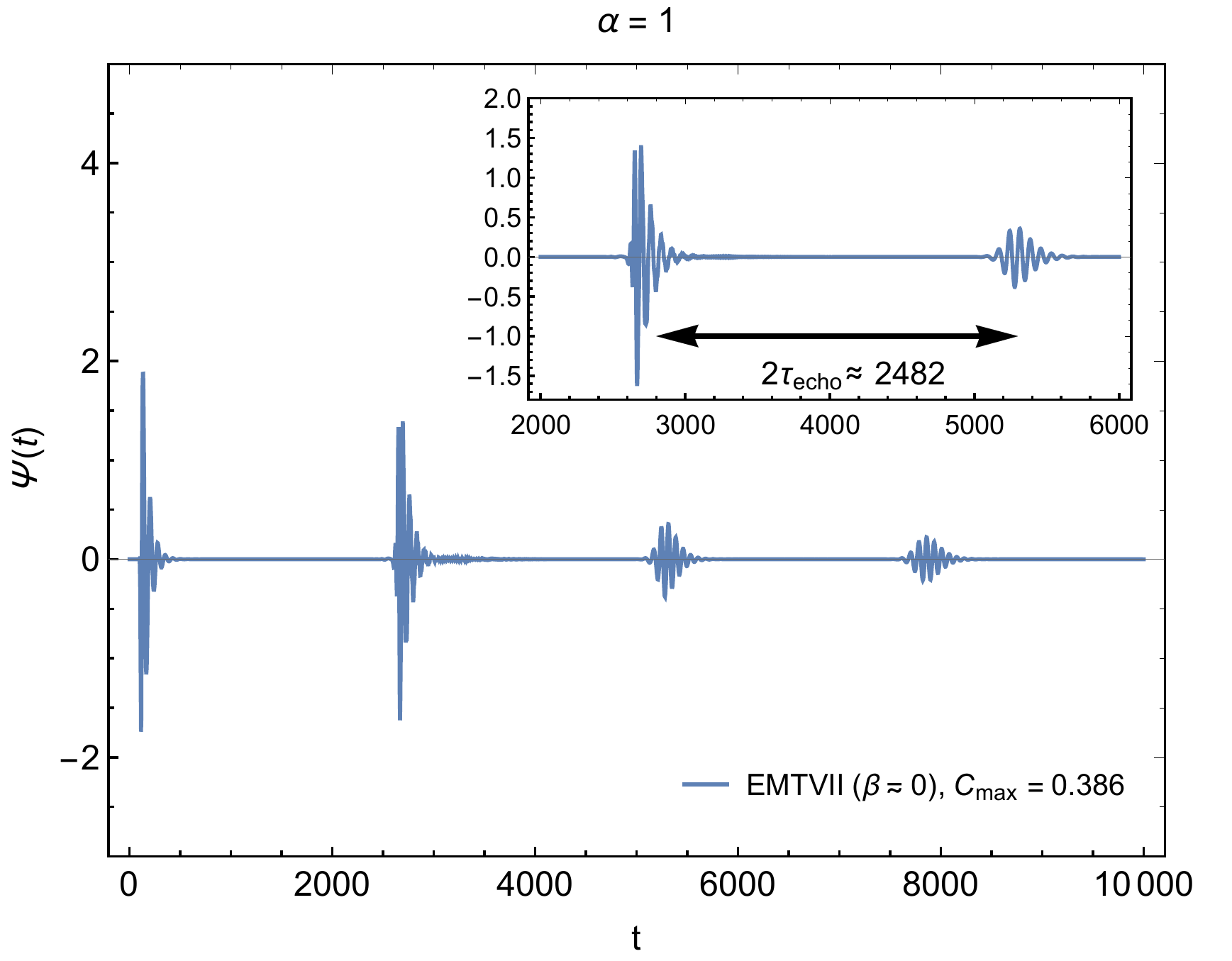}\\
	\includegraphics[width=0.45\linewidth]{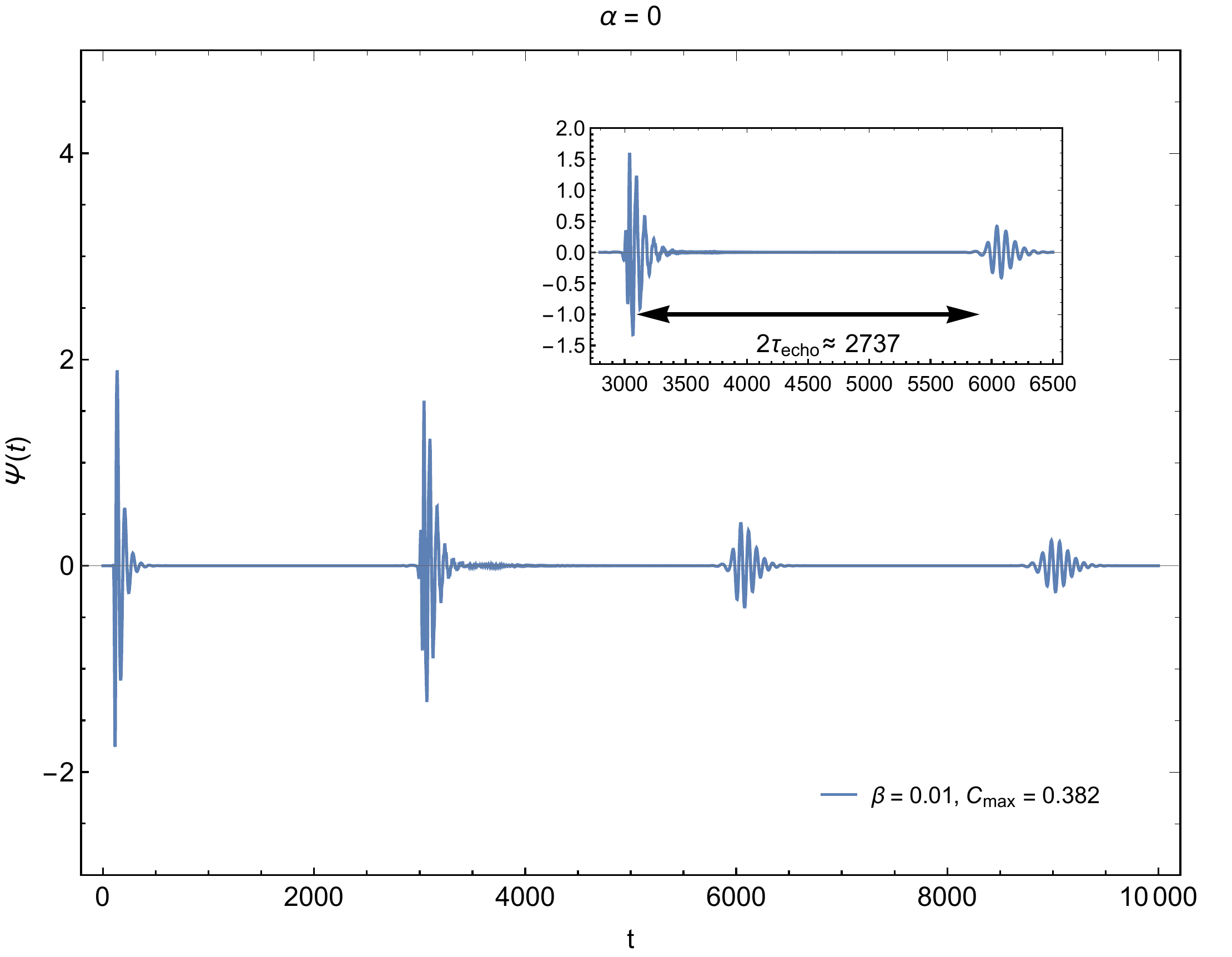}
       \includegraphics[width=0.45\linewidth]{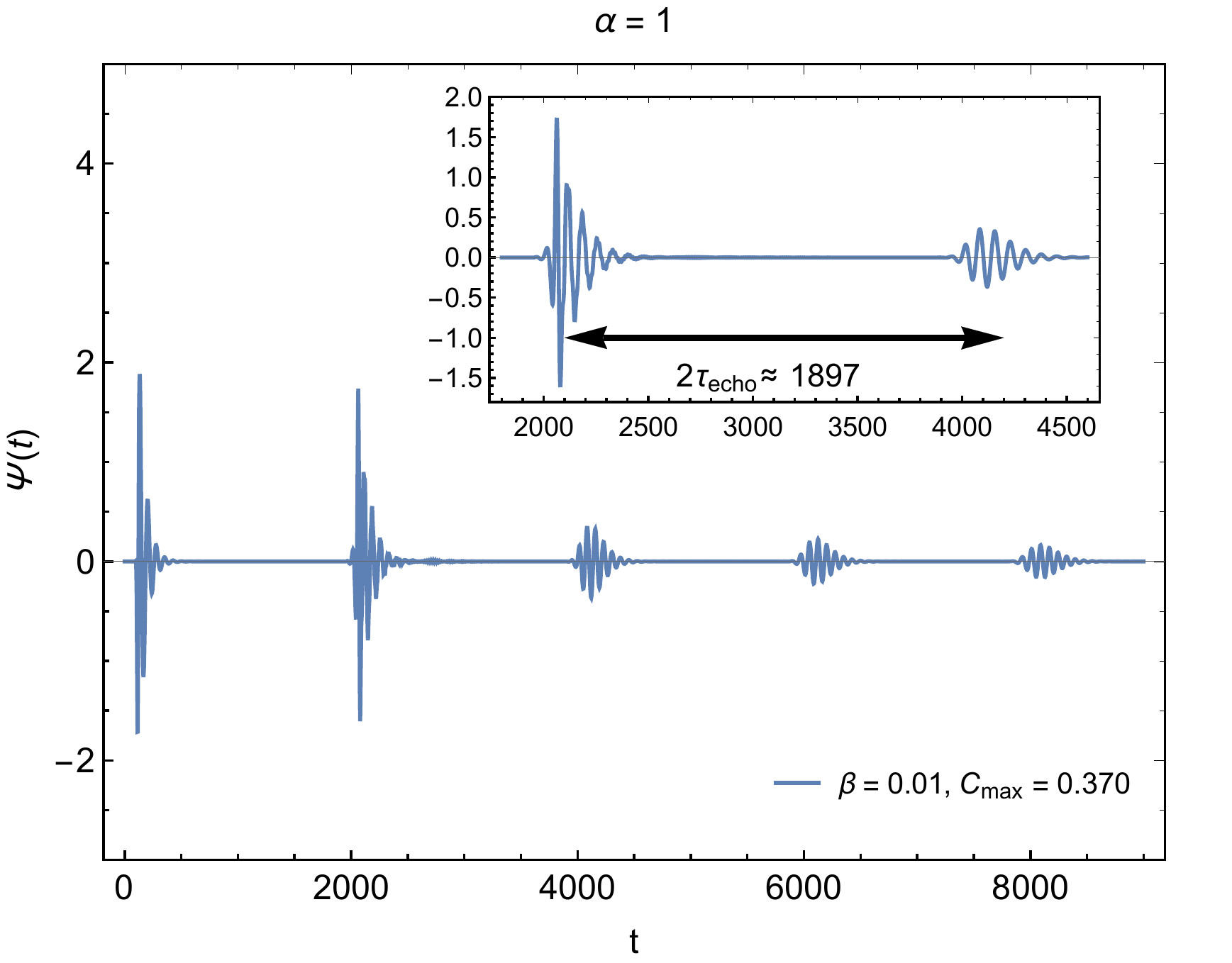}\\
	\includegraphics[width=0.45\linewidth]{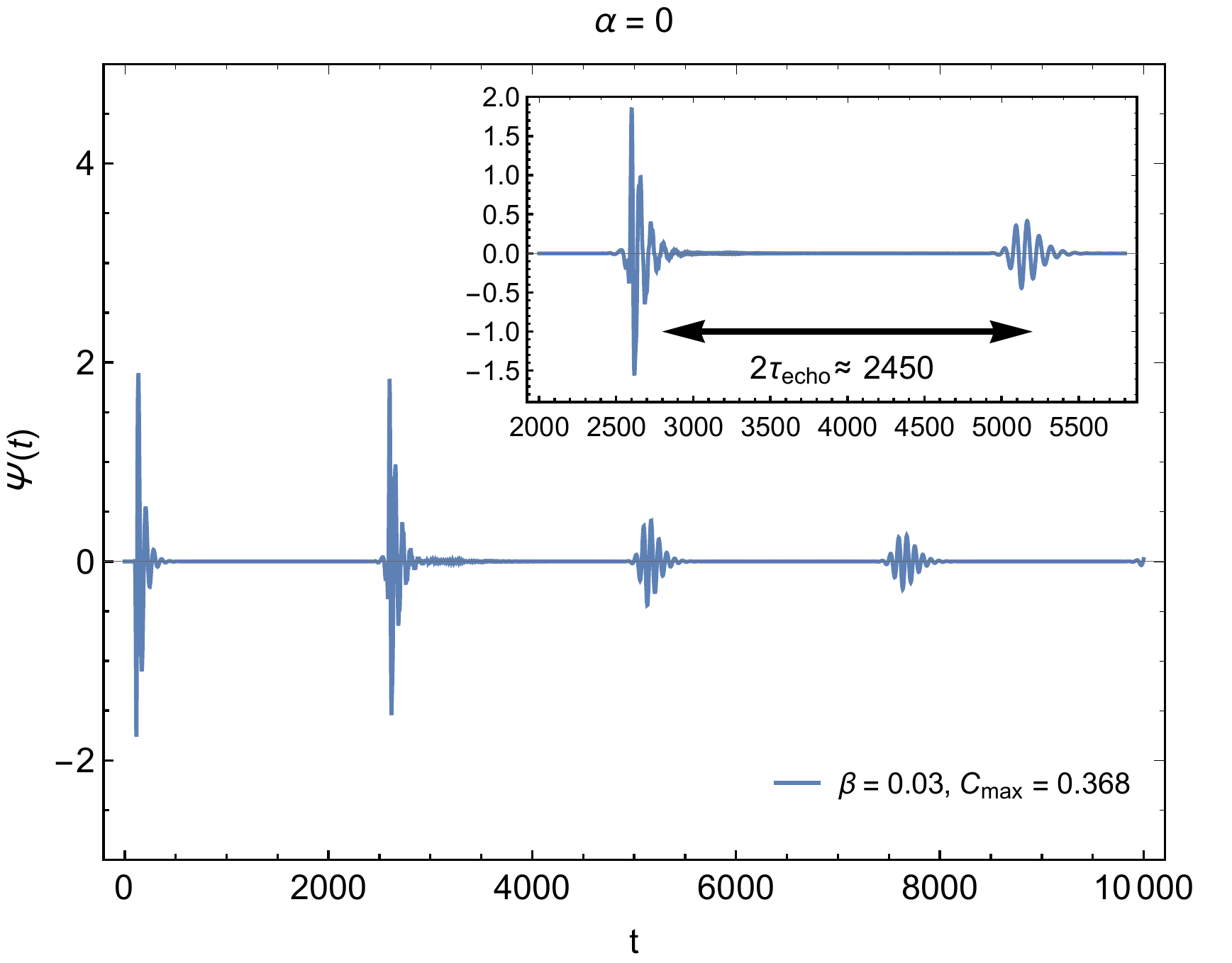}
      \includegraphics[width=0.45\linewidth]{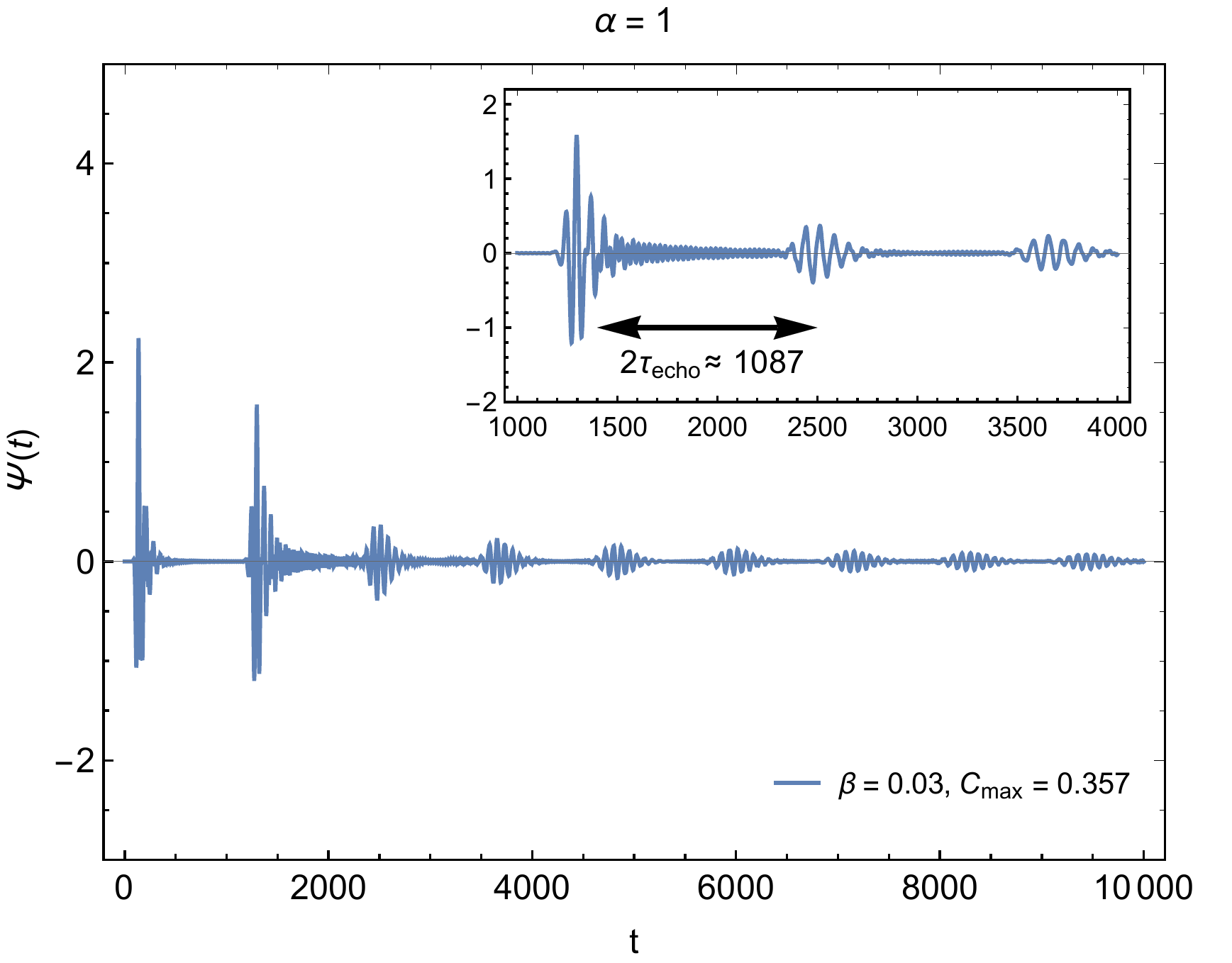}\\
	\caption{Eigen solution of the time-dependent Regge-Wheeler equation with respect to time for EMTVII~(upper) and NEMTVII~(middle and lower) model for $ \alpha=0,1$. The horizonless object in the ultracompact region produces the train echoes in both models only when DEC and causality condition are violated.}
	\label{fig:9}
\end{figure}

In this section, we numerically analyze the gravitational echoes from the perturbed ultracompact star by solving the time-dependent partial differential equation in Eqn. \eqref{pdp}. Since the highest differential operator in the equation is of the second order in $t$ and $ r_{*}, $ the solution requires four conditions. 
\begin{eqnarray}
\textrm{I.}&&~\Psi(r_{*},0)~~~~~~~~~~=~0,\nonumber\\
\textrm{II.}&&~\Psi(r_{*}^{c},t)~~~~~~~~~~=~0,\nonumber\\ 
\textrm{III.}&&~\frac{\partial \Psi(t,r_{*})}{\partial t}\bigg|_{t=0}~~~=~f(r_{*}),\nonumber\\
\textrm{IV.}&&~\frac{\partial \Psi(r_{*},t)}{\partial r}\bigg|_{r_{*}\rightarrow \infty} = -\frac{\partial \Psi(r_{*},t)}{\partial t}\bigg|_{r_{*}\rightarrow \infty}.
\end{eqnarray}
Condition I and III are the initial data and II and IV are the boundary conditions. Both initial data denotes the post-merger phase with an initial Gaussian pulse centered at $ r_{*}=r_{g} $ and with spread $ \sigma $; $ f(r_{*})=e^{-\frac{(r_{*}-r_{g})^2}{\sigma^2}} $ \cite{Urbano:2018nrs,Cardoso:2016oxy}. III is the condition for the star to be regular at the center, where $ r_{*}^{c}(r)=r_{*}(0). $ Lastly, condition IV relates outgoing waves at spatial infinity \cite{Urbano:2018nrs}. After evaluating numerically, the solution of the differential equation depends on the time and tortoise radius, $ \Psi(r_{*},t). $ We will focus on the time evolution of the signal from the EMTVII and NEMTVII model and present the lowest mode only. The signal can be numerically obtained by inserting the effective potential (as a function of tortoise coordinate $ r_{*} $) into the Eqn.~\eqref{pdp}. Since $ V_{\textrm{eff}} $ in the EMTVII model has wider interior, the signal needs more time to traverse the interior as shown in Fig. \ref{fig:9}. The effective potential of compact object with compactness consistent with the causal condition does not exhibit any well in the interior, and thus such a star does not produce the gravitational echoes even in the NEMTVII model.

\begin{figure}[h!]
	\centering
	\includegraphics[width=0.6\linewidth]{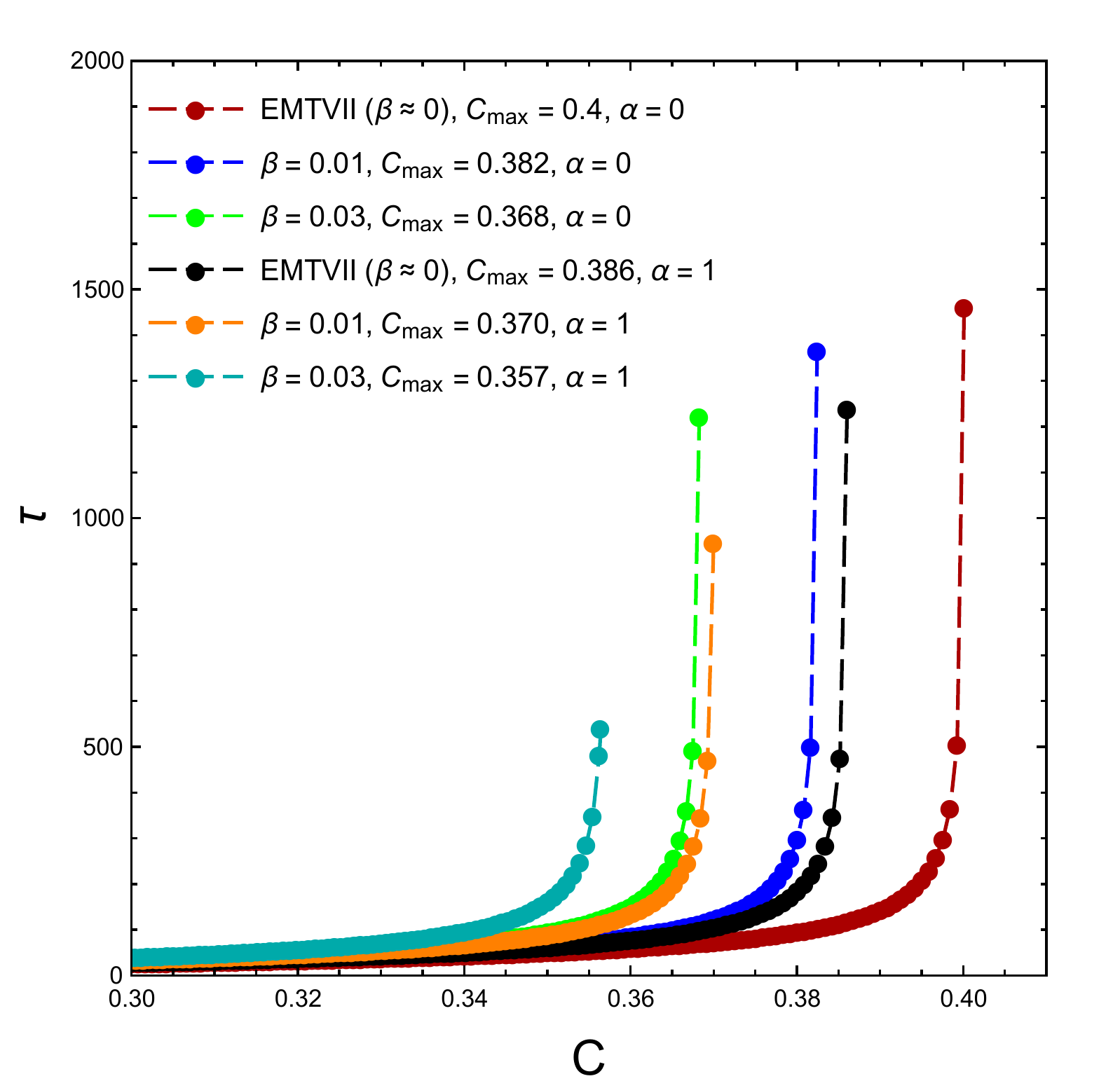}
	\caption{The echo time $ \tau $ versus compactness $ \mathcal{C} $ of the EMTVII and NEMTVII star. When the compactness approaches maximum, the echo time saturates to a certain bound.}
	\label{fig:10}
\end{figure}
Echo time is defined as the time needed for the waves to propagate from center of the star to the photon sphere radius \cite{Mannarelli:2018pjb,Pani:2018flj}. The echo time can be written as 
\begin{equation}
\tau_{echo} = \int_{0}^{3M} \sqrt{-\frac{g_{rr}}{g_{tt}}} ~dr.
\end{equation}
In our case, the solution is obtained numerically. The echo time is plotted by varying the compactness as shown in Fig.~\ref{fig:10}. In the EMTVII model, as denoted by the black solid line, the limiting compactness where $\tau \to \infty$ is far below the Buchdahl limit since its compactness is bounded in order to obtain finite pressure profile. The maximum compactness a star can reach is $ \mathcal{C}_{max}= 0.4 $ for $ \alpha=0 $. When we include the nonlocal effect, the star surface is smeared and the compactness decreases. Thus, the echo time curve shifts to lower values. The echo time corresponds to the propagation time from the star center to the light ring as depicted in Fig.~\ref{fig:9}. The ``$2\tau_{\textrm{echo}}$" denotes the time wave travels from light ring to the center and then back to the light ring.

On the other hand, the maximum compactness at which DEC is still minimally satisfied appears at $ \alpha=0 $, $ \beta=0.01 $, and $\mathcal{C} = 0.324$ (but causality condition, $c_{s}\leq 1$, is violated) for NEMTVII model. As shown in Fig.~\ref{fig:6.5}, the effective potential well almost disappeared and there is no echoes. Notably, this compactness is smaller than the corresponding value from EMTVII model where $\alpha=0,~\mathcal{C}=0.34$, just above the lower limit of the ultracompact object regime.


\section{Conclusions and Discussions}  \label{SectV}

Exact Modified Tolman VII (EMTVII), an extension of Tolman VII (TVII) with the additional free parameter, appears as a solution of realistic candidate EoS of NS. The numerical solution \cite{Posada:2022lij} shows that the additional free parameter, $\alpha$ is adequate to increase both the maximum compactness and the causal limit. If the compactness of the compact stars takes the value in the ultracompact regime, the gravitational perturbation can be more viable.

We introduce the nonlocal gravity contribution to the EMTVII solution (NEMTVII), which could provide the interior structure of the compact object in the ultracompact regime ($ 0.33 \leq \mathcal{C} \leq 0.44 $). In the Einstein field equation, the nonlocality only couples to the matter sector and therefore affects only the density and pressure. Notably, nonlocality smears out surface of the star and effectively reduce its compactness.

It is found that the maximum compactness for NEMTVII model is reduced to $ \mathcal{C}_{max}=0.382 $ for $ \beta=0.01 $ and $ \alpha=0 $ whether for EMTVII model the maximum compactness is $ \mathcal{C}_{max}=0.4 $ and $ \alpha=0 $. We further explore two important aspects of the gravitational perturbations: the quasinormal modes and the time-dependent gravitational wave echoes. By considering the effective potential in the tortoise coordinate, we found that the interior of the star in NEMTVII model has smaller well. As a consequence, the EMTVII model can excite more trapped modes than the nonlocal one. To study the echoes, gravitational waves solution of the time dependent Regge-Wheeler equation can be evaluated with appropriate boundary conditions. Each of the model exhibits train echoes from the incoming waves that enter the light ring barrier, propagate to the center and come back again to the light ring. The NEMTVII object has shorter echo time than the EMTVII counterpart due to the smaller potential well in the interior of the object.

For compact object in NEMTVII mass model with maximum compactness saturating causality condition such as NS with $ \mathcal{C}=0.2667,~\beta=0.01,~\alpha=1.4 $~(see Table~\ref{tab2}), the potential well and the light ring do not exist since the compactness is below the ultracompact regime. Notably this star has $M=2.14 M_{\odot}, R=11.835$ km, the star profile very close to the NS with multiquark core and stiff nuclear crust studied in Ref.~\cite{Pinkanjanarod:2020mgi,Pinkanjanarod:2021qto,Burikham:2021xpn} and the multiquark star studied in Ref.~\cite{Ponglertsakul:2022vni}. Our results in Fig.~\ref{figMR} explore certain classes of physical EoS that the EMTVII and NEMTVII parameter space could cover at low compactness where causality condition is {\t saturating} within the star.

It is also interesting to investigate how the NEMTVII model fit into observational constraints by considering the tidal deformability parametrized by Love number $ k_{2} $ \cite{Hinderer:2007mb}. There are literatures discussing the tidal deformation in Tolman VII and EMTVII~\cite{Posada:2022lij,Yagi:2013bca,Yagi:2013awa,Doneva:2017jop,Jiang:2020uvb}, we leave this for our further investigations.


\acknowledgments
We thank to anonymous reviewer for the technical comment. B.N. Jayawiguna is supported by the Second Century Fund (C2F), Chulalongkorn University, Thailand. P.B. is supported in part by National Research Council of Thailand~(NRCT) and Chulalongkorn University under Grant N42A660500. This research has received funding support from the NSRF via the Program Management Unit for Human Resources \& Institutional Development, Research and Innovation~[grant number B39G660025].


\section{Appendix} 

\subsection{Maximum Compactness ($ \mathcal{C}_{max} $)}   \label{AppA}

The maximum compactness is presented in Table~\ref{tab1}.

	\renewcommand{\arraystretch}{0.59} 
\begin{table}  
	\caption{Nonlocal parameter $ \beta $, new star radius and maximum compactness for $ \beta = 10^{-12},~0.01,~0.03.$}
	$
	\begin{array}{|c | c | c | c |}
	\hline & \textrm{EMTVII} &   \\ \hline
	\hline \alpha & R_{N}~({\rm km})  & \mathcal{C}_{\textrm{max}} \\ \hline
	0& 11.4 & 0.400  \\
	0.1& 11.4 & 0.399 \\
	0.2& 11.4 & 0.398 \\
	0.3& 11.4 & 0.397  \\
	0.4& 11.4 & 0.395  \\
	0.5& 11.4 & 0.394  \\
	0.6& 11.4 & 0.393  \\
	0.7& 11.4 & 0.391  \\
	0.8& 11.4 & 0.390  \\
	0.9& 11.4 & 0.388  \\
	1& 11.4 & 0.386 \\
	1.1& 11.4 & 0.383  \\
	1.2& 11.4 & 0.381  \\
	1.3& 11.4 & 0.378  \\
	1.4& 11.4 & 0.374  \\
	1.5& 11.4 & 0.371  \\
	1.6& 11.4 & 0.366  \\
	1.7& 11.4 & 0.362  \\
	1.8& 11.4 & 0.356  \\
	1.9& 11.4 & 0.349  \\
	2& 11.4 & 0.341  \\
	\hline
	\end{array}
	\hspace*{\fill}
	\begin{array}{|c | c | c | c |}
	\hline & \beta=0.01 &   \\ \hline
	\hline
	\alpha & R_{N}~({\rm km})  & \mathcal{C}_{\textrm{max}} \\ \hline
	0&11.933& 0.382  \\
	0.1& 11.930 & 0.381 \\
	0.2& 11.928 & 0.380 \\
	0.3& 11.925 & 0.379  \\
	0.4& 11.922 & 0.378  \\
	0.5& 11.919 & 0.377  \\
	0.6& 11.916 & 0.376  \\
	0.7& 11.912 & 0.375  \\
	0.8& 11.908 & 0.373  \\
	0.9& 11.90 & 0.371  \\
	1& 11.897 & 0.370  \\
	1.1& 11.891 & 0.368  \\
	1.2& 11.88 & 0.365  \\
	1.3& 11.87 & 0.363  \\
	1.4& 11.865 & 0.360  \\
	1.5& 11.852 & 0.357  \\
	1.6& 11.836 & 0.353  \\
	1.7& 11.815 & 0.349  \\
	1.8& 11.784 & 0.344  \\
	1.9& 11.734 & 0.339  \\
	2& 11.60 & 0.335  \\
	\hline
	\end{array}
	\hspace*{\fill}
	\begin{array}{|c | c | c | c |}
	\hline & \beta = 0.03 &   \\ \hline
	\hline
	\alpha & R_{N} ({\rm km})  & \mathcal{C}_{\textrm{max}} \\ \hline
	0&12.406& 0.368  \\
	0.1& 12.402 & 0.367 \\
	0.2& 12.398 & 0.366 \\
	0.3& 12.394 & 0.365  \\
	0.4& 12.389 & 0.364 \\
	0.5& 12.384 & 0.363  \\
	0.6& 12.378 & 0.362  \\
	0.7& 12.372 & 0.361  \\
	0.8& 12.364 & 0.360  \\
	0.9& 12.356 & 0.358  \\
	1& 12.347 & 0.357  \\
	1.1& 12.337 & 0.355  \\
	1.2& 12.325 & 0.353  \\
	1.3& 12.311 & 0.350  \\
	1.4& 12.294 & 0.348  \\
	1.5& 12.273 & 0.345  \\
	1.6& 12.247 & 0.342  \\
	1.7& 12.213 & 0.338  \\
	1.8& 12.166 & 0.334  \\
	1.9& 12.092 & 0.330  \\
	2& 1.936 & 0.326 \\
	\hline
	\end{array}   
	$ 
	\label{tab1}
\end{table}

\subsection{Compactness within the Causal Condition, $ c_{s}\leq 1 $ }  \label{AppB}

The compactness within causality condition is presented in Table~\ref{tab2}.

\begin{table}[h!]  
	\caption{Compactness within the causal condition ($ c_{s}\leq1 $).}
	$
	\begin{array}{|c | c | c | c |}
	\hline & \textrm{EMTVII} &   \\ \hline
	\hline \alpha & R_{N}~({\rm km})  & \mathcal{C}_{\textrm{max}} \\ \hline
	0& 11.4 & 10^{-5}  \\
	0.1& 11.4 & 0.084 \\
	0.2& 11.4 & 0.144 \\
	0.3& 11.4 & 0.179  \\
	0.4& 11.4 & 0.208  \\
	0.5& 11.4 & 0.226  \\
	0.6& 11.4 & 0.240  \\
	0.7& 11.4 & 0.250  \\
	0.8& 11.4 & 0.259  \\
	0.9& 11.4 & 0.265  \\
	1& 11.4 & 0.269 \\
	1.1& 11.4 & 0.273  \\
	1.2& 11.4 & 0.275  \\
	1.3& 11.4 & 0.2765  \\
	1.4& 11.4 & 0.276  \\
	1.5& 11.4 & 0.275  \\
	1.6& 11.4 & 0.274  \\
	1.7& 11.4 & 0.272  \\
	1.8& 11.4 & 0.268  \\
	1.9& 11.4 & 0.263  \\
	2& 11.4 & 0.258  \\
	\hline
	\end{array}
	\hspace*{\fill}
	\begin{array}{|c | c | c | c |}
	\hline & \beta=0.01 &   \\ \hline
	\hline
	\alpha & R_{N}~({\rm km})  & \mathcal{C}_{\textrm{max}} \\ \hline
	0  &11.438& 0.001  \\
	0.1& 11.782 & 0.086 \\
	0.2& 11.828 & 0.139 \\
	0.3& 11.848 & 0.175  \\
	0.4& 11.858 & 0.199 \\
	0.5& 11.864 & 0.217  \\
	0.6& 11.867 & 0.230  \\
	0.7& 11.868 & 0.240  \\
	0.8& 11.867 & 0.249  \\
	0.9& 11.865 & 0.254  \\
	1  & 11.862 & 0.259  \\
	1.1& 11.857 & 0.262  \\
	1.2& 11.851 & 0.265  \\
	1.3& 11.844 & 0.2664  \\
	1.4& 11.835 & 0.2667  \\
	1.5& 11.823 & 0.2664  \\
	1.6& 11.807 & 0.265  \\
	1.7& 11.786 & 0.263  \\
	1.8& 11.756 & 0.260  \\
	1.9& 11.706 & 0.257  \\
	2  & 11.578 & 0.254 \\
	\hline
	\end{array}
		\hspace*{\fill}
	\begin{array}{|c | c | c | c |}
	\hline & \beta = 0.2 &   \\ \hline
	\hline
	\alpha & R_{N} ({\rm km})  & \mathcal{C}_{\textrm{max}} \\ \hline
	0&13.218& 0.028  \\
	0.1& 13.733 & 0.090 \\
	0.2& 13.885 & 0.128 \\
	0.3& 13.957 & 0.154  \\
	0.4& 14 & 0.173 \\
	0.5& 14.024 & 0.188  \\
	0.6& 14.038 & 0.198  \\
	0.7& 14.044 & 0.207  \\
	0.8& 14.041 & 0.213  \\
	0.9& 14.03 & 0.218  \\
	1& 14.024 & 0.221  \\
	1.1& 14.012 & 0.225  \\
	1.2& 13.991 & 0.227  \\
	1.3& 13.968 & 0.2291  \\
	1.4& 13.935 & 0.2290  \\
	1.5& 13.896 & 0.2294  \\
	1.6& 13.847 & 0.2290  \\
	1.7& 13.784 & 0.227  \\
	1.8& 13.700 & 0.226  \\
	1.9& 13.580 & 0.224  \\
	2& 13.395 & 0.222 \\
	\hline
	\end{array}   
	$ 
	
	\label{tab2}
\end{table}

\newpage
\subsection{Table for QNM}    \label{AppC}

The QNMs are presented in Table~\ref{tab3}, \ref{tab4}. 
 
\begin{table}\centering  
	\ra{0.68}
	\caption{The lowest-mode QNM for EMTVII ($ \alpha=0~\textrm{and} ~ \mathcal{C}_{max}=0.4 $) and NEMTVII ($ \beta=0.03,~\alpha=0,~\mathcal{C}_{max}=0.368 $).} 

	\begin{tabular}{@{}rrrrcrrrcrrr@{}}\toprule
		~~~~~~~~~$ l=2, $  & \multicolumn{3}{c}{$ \textrm{\textbf{EMTVII}}$} ~~~~~~~~~~~& & \multicolumn{3}{c}{$ \textrm{\textbf{NEMTVII}} $}~~~~~ & & \multicolumn{3}{c}{}  \\
		\cmidrule{2-4} \cmidrule{6-8} \cmidrule{10-12}
		$ n $	~~~~~& Re$ (\omega_{n}) $ 	~~~~~~& $ \big|\textrm{Im}(\omega_{n})\big| $ ~~~~~~~~~& && Re$ (\omega_{n}) $ 	  ~~~~& $ \big|\textrm{Im}(\omega_{n})\big| $	~~~~& &&   & \\ \midrule
		$0$  ~~~~~& 0.0067 	~~~~~~& $ 5.16 \times 10^{-15}$ ~~~~~~~& && 0.0081   ~~~~~& $ 2.91 \times 10^{-14} $ &  \\
		$1$  ~~~~~& 0.0112	~~~~~~& $ 3.69 \times 10^{-13}$	~~~~~~~& && 0.0134   ~~~~~& $ 2.14 \times 10^{-12} $	&  \\
		$2$  ~~~~~& 0.0151	~~~~~~& $ 4.49 \times 10^{-12}$	~~~~~~~& && 0.0180   ~~~~~& $ 2.68 \times 10^{-10} $	&  \\
		$3$  ~~~~~& 0.0185	~~~~~~& $ 2.63 \times 10^{-11}$	~~~~~~~& && 0.0221   ~~~~~& $ 1.61 \times 10^{-10} $	&  \\
		$4$  ~~~~~& 0.0217	~~~~~~& $ 1.04 \times 10^{-10}$	~~~~~~~& && 0.0259   ~~~~~& $ 6.60 \times 10^{-10} $ &  \\
		$5$  ~~~~~& 0.0247	~~~~~~& $ 3.23 \times 10^{-10}$	~~~~~~~& && 0.0295   ~~~~~& $ 2.11 \times 10^{-9} $	&  \\
		$6$  ~~~~~& 0.0275	~~~~~~& $ 8.51 \times 10^{-10}$ ~~~~~~~& && 0.0328   ~~~~~& $ 5.71 \times 10^{-9} $	&  \\
		$7$  ~~~~~& 0.0302	~~~~~~& $ 1.98 \times 10^{-9}$ ~~~~~~~& && 0.0360   ~~~~~& $ 1.37 \times 10^{-8} $	&  \\
		$8$  ~~~~~& 0.0328	~~~~~~& $ 4.22 \times 10^{-9}$ ~~~~~~~& && 0.0391   ~~~~~& $ 2.00 \times 10^{-8} $	&  \\
		$9$  ~~~~~& 0.0352	~~~~~~& $ 8.35 \times 10^{-9} $	~~~~~~~& && 0.0421   ~~~~~& $ 6.12 \times 10^{-8} $	&  \\
		$10$ ~~~~~& 0.0376	~~~~~~& $ 1.56 \times 10^{-8} $	~~~~~~~& && 0.0449   ~~~~~& $ 1.17 \times 10^{-7} $	&  \\
		$11$ ~~~~~& 0.0399	~~~~~~& $ 2.78 \times 10^{-8} $	~~~~~~~& && 0.0477   ~~~~~& $ 2.16 \times 10^{-7} $	&  \\
		$12$ ~~~~~& 0.0422	~~~~~~& $ 4.77 \times 10^{-8} $	~~~~~~~& && 0.0504   ~~~~~& $ 3.81 \times 10^{-7} $	&  \\
		$13$ ~~~~~& 0.0444	~~~~~~& $ 7.90 \times 10^{-8} $	~~~~~~~& && 0.0530   ~~~~~& $ 6.49 \times 10^{-7} $	&  \\
		$14$ ~~~~~& 0.0465	~~~~~~& $ 1.27 \times 10^{-7} $	~~~~~~~& && 0.0556   ~~~~~& $ 1.07 \times 10^{-6} $	&  \\
		$15$ ~~~~~& 0.0486	~~~~~~& $ 1.99 \times 10^{-7} $	~~~~~~~& && 0.0581   ~~~~~& $ 1.73 \times 10^{-6} $	&  \\
		$16$ ~~~~~& 0.0507	~~~~~~& $ 3.06 \times 10^{-7} $	~~~~~~~& && 0.0605   ~~~~~& $ 2.74 \times 10^{-6} $	&  \\
		$17$ ~~~~~& 0.0527	~~~~~~& $ 4.61 \times 10^{-7} $	~~~~~~~& && 0.0629   ~~~~~& $ 4.24 \times 10^{-6} $	&  \\
		$18$ ~~~~~& 0.0547	~~~~~~& $ 6.28 \times 10^{-7} $	~~~~~~~& && 0.0653   ~~~~~& $ 6.44 \times 10^{-6} $	&  \\
		$19$ ~~~~~& 0.0566	~~~~~~& $ 9.93 \times 10^{-7} $	~~~~~~~& && 0.0676   ~~~~~& $ 9.65 \times 10^{-6} $	&  \\
		$20$ ~~~~~& 0.0585	~~~~~~& $ 1.42 \times 10^{-6} $	~~~~~~~& && 0.0699   ~~~~~& $ 1.42 \times 10^{-5} $	&  \\
		$21$ ~~~~~& 0.0604	~~~~~~& $ 2.02 \times 10^{-6} $	~~~~~~~& && 0.0721   ~~~~~& $ 2.07 \times 10^{-5} $	&  \\
		$22$ ~~~~~& 0.0623	~~~~~~& $ 2.83 \times 10^{-6} $	~~~~~~~& && 0.0743   ~~~~~& $ 2.98 \times 10^{-5} $	&  \\
		$23$ ~~~~~& 0.0641	~~~~~~& $ 3.92 \times 10^{-6} $	~~~~~~~& && 0.0765   ~~~~~& $ 4.24 \times 10^{-5} $	&  \\
		$24$ ~~~~~& 0.0659	~~~~~~& $ 5.38 \times 10^{-6} $	~~~~~~~& && 0.0786   ~~~~~& $ 5.98 \times 10^{-5} $	&  \\
		$25$ ~~~~~& 0.0677	~~~~~~& $ 7.32 \times 10^{-7} $	~~~~~~~& && 0.0808   ~~~~~& $ 8.35 \times 10^{-5} $	&  \\
		$26$ ~~~~~& 0.0694	~~~~~~& $ 9.88 \times 10^{-6} $	~~~~~~~& && 0.0828   ~~~~~& $ 1.15 \times 10^{-4} $	&  \\
		$27$ ~~~~~& 0.0711	~~~~~~& $ 1.32 \times 10^{-5} $	~~~~~~~& && 0.0849    ~~~~~& $ 1.58 \times 10^{-4} $	&  \\
		$28$ ~~~~~& 0.0728	~~~~~~& $ 1.75 \times 10^{-5} $	~~~~~~~& && 	     ~~~~~& 	&  \\
		$29$ ~~~~~& 0.0745	~~~~~~& $ 2.31 \times 10^{-5} $	~~~~~~~& &&    		 ~~~~~&		&  \\
		$30$ ~~~~~& 0.0762	~~~~~~& $ 3.03 \times 10^{-5} $	~~~~~~~& &&          ~~~~~&     &  \\
		$31$ ~~~~~& 0.0778	~~~~~~& $ 3.95 \times 10^{-5} $	~~~~~~~& &&          ~~~~~& 	&  \\
		$32$ ~~~~~& 0.0795	~~~~~~& $ 5.12 \times 10^{-5} $	~~~~~~~& &&          ~~~~~& 	&  \\
		$33$ ~~~~~& 0.0811	~~~~~~& $ 6.61 \times 10^{-5} $	~~~~~~~& &&          ~~~~~& 	&  \\
		$34$ ~~~~~& 0.0827	~~~~~~& $ 8.48 \times 10^{-5} $	~~~~~~~& &&          ~~~~~& 	&  \\
		$35$ ~~~~~& 0.0843	~~~~~~& $ 1.08 \times 10^{-4} $	~~~~~~~& &&          ~~~~~&	    &  \\
		\bottomrule
	\end{tabular}
 \label{tab3}
\end{table}

\begin{table}\centering   
	\ra{0.7}
	\caption{QNMs with $ l=3 $ for EMTVII ($ \alpha=0~\textrm{and}~ \mathcal{C}_{max}=0.40 $) and NEMTVII ($ \beta=0.03,~\alpha=0,~\mathcal{C}_{max}=0.368 $).}  
	\begin{tabular}{@{}rrrrcrrrcrrr@{}}\toprule
		~~~~~~~~~$ l=3, $  & \multicolumn{3}{c}{$ \textrm{\textbf{EMTVII}}$} ~~~~~~~~~~~& & \multicolumn{3}{c}{$ \textrm{\textbf{NEMTVII}} $}~~~~~ & & \multicolumn{3}{c}{}  \\
		\cmidrule{2-4} \cmidrule{6-8} \cmidrule{10-12}
		$ n $	~~~~~& Re$ (\omega_{n}) $ 	~~~~~~& $ \big|\textrm{Im}(\omega_{n})\big| $ ~~~~~~~~~& && Re$ (\omega_{n}) $ 	  ~~~~& $ \big|\textrm{Im}(\omega_{n})\big| $	~~~~& &&   & \\ \midrule
		$0$  ~~~~~& 0.0090 	~~~~~~& $ 1.36 \times 10^{-21}$ ~~~~~~~& && 0.0107   ~~~~~& $ 1.63 \times 10^{-20} $ &  \\
		$1$  ~~~~~& 0.0138	~~~~~~& $ 3.56 \times 10^{-19}$	~~~~~~~& && 0.0163   ~~~~~& $ 4.19 \times 10^{-18} $	&  \\
		$2$  ~~~~~& 0.0181	~~~~~~& $ 1.23 \times 10^{-17}$	~~~~~~~& && 0.0215   ~~~~~& $ 1.46 \times 10^{-16} $	&  \\
		$3$  ~~~~~& 0.0220	~~~~~~& $ 1.61 \times 10^{-16}$	~~~~~~~& && 0.0261   ~~~~~& $ 1.94 \times 10^{-15} $	&  \\
		$4$  ~~~~~& 0.0257	~~~~~~& $ 1.20 \times 10^{-15}$	~~~~~~~& && 0.0304   ~~~~~& $ 1.49 \times 10^{-14} $	&  \\
		$5$  ~~~~~& 0.0291	~~~~~~& $ 6.36 \times 10^{-15}$	~~~~~~~& && 0.0345   ~~~~~& $ 8.08 \times 10^{-14} $	&  \\
		$6$  ~~~~~& 0.0324	~~~~~~& $ 2.62 \times 10^{-14}$	~~~~~~~& && 0.0384   ~~~~~& $ 3.42 \times 10^{-13} $	&  \\
		$7$  ~~~~~& 0.0355	~~~~~~& $ 9.06 \times 10^{-14}$	~~~~~~~& && 0.0420   ~~~~~& $ 1.21 \times 10^{-12} $	&  \\
		$8$  ~~~~~& 0.0385	~~~~~~& $ 2.72 \times 10^{-13}$	~~~~~~~& && 0.0455   ~~~~~& $ 3.74 \times 10^{-12} $ &  \\
		$9$  ~~~~~& 0.0413	~~~~~~& $ 7.36 \times 10^{-13}$	~~~~~~~& && 0.0489   ~~~~~& $ 1.03 \times 10^{-11} $	&  \\
		$10$  ~~~~~& 0.0441	~~~~~~& $ 1.82 \times 10^{-12}$ ~~~~~~~& && 0.0522   ~~~~~& $ 2.64 \times 10^{-11} $	&  \\
		$11$  ~~~~~& 0.0468	~~~~~~& $ 4.20 \times 10^{-12}$ ~~~~~~~& && 0.0554   ~~~~~& $ 6.26 \times 10^{-11} $	&  \\
		$12$  ~~~~~& 0.0494	~~~~~~& $ 9.13 \times 10^{-12}$ ~~~~~~~& && 0.0585   ~~~~~& $ 1.39 \times 10^{-10} $	&  \\
		$13$  ~~~~~& 0.0520	~~~~~~& $ 1.88 \times 10^{-11} $	~~~~~~~& && 0.0615   ~~~~~& $ 2.95 \times 10^{-10} $	&  \\
		$14$ ~~~~~& 0.0545	~~~~~~& $ 3.71 \times 10^{-11} $	~~~~~~~& && 0.0645   ~~~~~& $ 6.00 \times 10^{-10} $	&  \\
		$15$ ~~~~~& 0.0569	~~~~~~& $ 7.06 \times 10^{-11} $	~~~~~~~& && 0.0674   ~~~~~& $ 1.17 \times 10^{-9} $	&  \\
		$16$ ~~~~~& 0.0593	~~~~~~& $ 1.29 \times 10^{-10} $	~~~~~~~& && 0.0702   ~~~~~& $ 2.20 \times 10^{-9} $	&  \\
		$17$ ~~~~~& 0.0617	~~~~~~& $ 2.31 \times 10^{-10} $	~~~~~~~& && 0.0730   ~~~~~& $ 4.04 \times 10^{-9} $	&  \\
		$18$ ~~~~~& 0.0640	~~~~~~& $ 4.01 \times 10^{-10} $	~~~~~~~& && 0.0757   ~~~~~& $ 7.20 \times 10^{-9} $	&  \\
		$19$ ~~~~~& 0.0662	~~~~~~& $ 6.81 \times 10^{-10} $	~~~~~~~& && 0.0784   ~~~~~& $ 1.25 \times 10^{-8} $	&  \\
		$20$ ~~~~~& 0.0685	~~~~~~& $ 1.13 \times 10^{-9} $	~~~~~~~& && 0.0810   ~~~~~& $ 2.13 \times 10^{-8} $	&  \\
		$21$ ~~~~~& 0.0706	~~~~~~& $ 1.83 \times 10^{-9} $	~~~~~~~& && 0.0836   ~~~~~& $ 3.56 \times 10^{-8} $	&  \\
		$22$ ~~~~~& 0.0728	~~~~~~& $ 2.94 \times 10^{-9} $	~~~~~~~& && 0.0861   ~~~~~& $ 5.85 \times 10^{-8} $	&  \\
		$23$ ~~~~~& 0.0749	~~~~~~& $ 4.62 \times 10^{-9} $	~~~~~~~& && 0.0887   ~~~~~& $ 9.45 \times 10^{-8} $	&  \\
		$24$ ~~~~~& 0.0770	~~~~~~& $ 7.16 \times 10^{-9} $	~~~~~~~& && 0.0911   ~~~~~& $ 1.50 \times 10^{-7} $	&  \\
		$25$ ~~~~~& 0.0791	~~~~~~& $ 1.09 \times 10^{-8} $	~~~~~~~& && 0.0936   ~~~~~& $ 2.35 \times 10^{-7} $	&  \\
		$26$ ~~~~~& 0.0811	~~~~~~& $ 1.65 \times 10^{-8} $	~~~~~~~& && 0.0960   ~~~~~& $ 3.64 \times 10^{-7} $	&  \\
		$27$ ~~~~~& 0.0831	~~~~~~& $ 2.46 \times 10^{-8} $	~~~~~~~& && 0.0984   ~~~~~& $ 5.58 \times 10^{-7} $	&  \\
		$28$ ~~~~~& 0.0851	~~~~~~& $ 3.63 \times 10^{-8} $	~~~~~~~& && 0.1007   ~~~~~& $ 8.44 \times 10^{-7} $	&  \\
		$29$ ~~~~~& 0.0871	~~~~~~& $ 5.30 \times 10^{-8} $	~~~~~~~& && 0.1031   ~~~~~& $ 1.26 \times 10^{-6} $	&  \\
		$30$ ~~~~~& 0.0890	~~~~~~& $ 7.67 \times 10^{-8} $	~~~~~~~& && 0.1054   ~~~~~& $ 1.87 \times 10^{-6} $	&  \\
		\bottomrule
	\end{tabular}
\label{tab4}
\end{table}
\begin{table}\centering   
	\ra{0.7}
	\begin{tabular}{@{}rrrrcrrrcrrr@{}}\toprule
		~~~~~~~~~$  $  & \multicolumn{3}{c}{$ \textrm{\textbf{EMTVII}}$} ~~~~~~~~~~~& & \multicolumn{3}{c}{$ \textrm{\textbf{NEMTVII}} $}~~~~~ & & \multicolumn{3}{c}{}  \\
		\cmidrule{2-4} \cmidrule{6-8} \cmidrule{10-12}
		$ n $	~~~~~& Re$ (\omega_{n}) $ 	~~~~~~& $ \big|\textrm{Im}(\omega_{n})\big| $ ~~~~~~~~~& && Re$ (\omega_{n}) $ 	  ~~~~& $ \big|\textrm{Im}(\omega_{n})\big| $	~~~~& &&   & \\ \midrule
		$31$ ~~~~~& 0.0910	~~~~~~& $ 1.10 \times 10^{-7} $	~~~~~~~& && 0.1076   ~~~~~& $ 2.75 \times 10^{-6} $	&  \\
		$32$ ~~~~~& 0.0929	~~~~~~& $ 1.56 \times 10^{-7} $	~~~~~~~& && 0.1099   ~~~~~& $ 4.01 \times 10^{-6} $	&  \\
		$33$ ~~~~~& 0.0947	~~~~~~& $ 2.20 \times 10^{-7} $	~~~~~~~& && 0.1121   ~~~~~& $ 5.80 \times 10^{-6} $ &  \\
		$34$ ~~~~~& 0.0966	~~~~~~& $ 3.08 \times 10^{-7} $	~~~~~~~& && 0.1143   ~~~~~& $ 8.31 \times 10^{-6} $	&  \\
		$35$ ~~~~~& 0.0985	~~~~~~& $ 4.28 \times 10^{-7} $	~~~~~~~& && 0.1165   ~~~~~& $ 1.18 \times 10^{-5} $	&  \\
		$36$ ~~~~~& 0.1003	~~~~~~& $ 5.91 \times 10^{-7} $	~~~~~~~& && 0.1187   ~~~~~& $ 1.67 \times 10^{-5} $	&  \\
		$37$ ~~~~~& 0.1021	~~~~~~& $ 8.11 \times 10^{-7} $	~~~~~~~& && 0.1208   ~~~~~& $ 2.35 \times 10^{-5} $	&  \\
		$38$ ~~~~~& 0.1039	~~~~~~& $ 1.10 \times 10^{-6} $	~~~~~~~& && 0.1230   ~~~~~& $ 3.29 \times 10^{-5} $	&  \\
		$39$ ~~~~~& 0.1057	~~~~~~& $ 1.49 \times 10^{-6} $	~~~~~~~& && 0.1251   ~~~~~& $ 4.56 \times 10^{-5} $	&  \\
		$40$ ~~~~~& 0.1075	~~~~~~& $ 2.02 \times 10^{-6} $	~~~~~~~& && 0.1272   ~~~~~& $ 6.31 \times 10^{-5} $	&  \\
		$41$ ~~~~~& 0.1092	~~~~~~& $ 2.70 \times 10^{-6} $	~~~~~~~& && 0.1292   ~~~~~& $ 8.67 \times 10^{-5} $	&  \\
		$42$ ~~~~~& 0.1110	~~~~~~& $ 3.61 \times 10^{-6} $	~~~~~~~& && 0.1313   ~~~~~& $ 1.18 \times 10^{-4} $	&  \\
		$43$ ~~~~~& 0.1127	~~~~~~& $ 4.80 \times 10^{-6} $	~~~~~~~& && 0.1333   ~~~~~& $ 1.61 \times 10^{-4} $	&  \\
		$44$ ~~~~~& 0.1144	~~~~~~& $ 6.35 \times 10^{-6} $~~~~~~~& &&    ~~~~~& 	&  \\
		$45$ ~~~~~& 0.1161	~~~~~~& $ 8.36 \times 10^{-6} $~~~~~~~& &&    ~~~~~& 	&  \\
		$46$ ~~~~~& 0.1178	~~~~~~& $ 1.09 \times 10^{-5} $~~~~~~~& &&    ~~~~~& 	&  \\
		$47$ ~~~~~& 0.1195	~~~~~~& $ 1.43 \times 10^{-5} $~~~~~~~& &&    ~~~~~& 	&  \\
		$48$ ~~~~~& 0.1211	~~~~~~& $ 1.86 \times 10^{-5} $~~~~~~~& &&    ~~~~~& 	&  \\
		$49$ ~~~~~& 0.1228	~~~~~~& $ 2.41 \times 10^{-5} $~~~~~~~& &&    ~~~~~& 	&  \\
		$50$ ~~~~~& 0.1244 	~~~~~~& $ 3.11 \times 10^{-5} $~~~~~~~& &&    ~~~~~& 	&  \\
		$51$ ~~~~~& 0.1261	~~~~~~& $ 4.01 \times 10^{-5} $~~~~~~~& &&    ~~~~~& 	&  \\
		$52$ ~~~~~& 0.1277	~~~~~~& $ 5.14 \times 10^{-5} $~~~~~~~& &&    ~~~~~& 	&  \\
		$53$ ~~~~~& 0.1293	~~~~~~& $ 6.58 \times 10^{-5} $~~~~~~~& &&    ~~~~~& 	&  \\
		$54$ ~~~~~& 0.1309	~~~~~~& $ 8.39 \times 10^{-5} $~~~~~~~& &&    ~~~~~& 	&  \\
		$55$ ~~~~~& 0.1325	~~~~~~& $ 1.06 \times 10^{-4} $~~~~~~~& &&    ~~~~~& 	&  \\
		\bottomrule
	\end{tabular}
\end{table}

\end{document}